\documentclass[a4paper,prl,twocolumn,preprintnumbers,floatfix,superscriptaddress,nofootinbib]{revtex4-2}
\usepackage[english]{babel}
\PassOptionsToPackage{utf8}{inputenc}
\usepackage{inputenc}
\PassOptionsToPackage{T1}{fontenc}
\usepackage{fontenc}
\usepackage{graphicx}
\usepackage{color}
\usepackage[separate-uncertainty=false]{siunitx}
\usepackage{amsmath}
\usepackage{mathtools}
\usepackage{braket}
\usepackage{hyperref}
\newcounter{supequation}
\makeatletter
\@addtoreset{equation}{supequation}
\makeatother
\newcounter{supfigure}
\makeatletter
\@addtoreset{figure}{supfigure}
\makeatother
\usepackage{xspace}
\newcommand{\al}{$^{27}\mathrm{Al}^+$\xspace}
\newcommand{\indium}{$^{115}\mathrm{In}^+$\xspace}
\newcommand{\ca}{$^{40}\mathrm{Ca}^+$\xspace}
\newcommand{\sr}{$^{88}\mathrm{Sr}^+$\xspace}
\newcommand{\lu}{$^{176}\mathrm{Lu}^+$\xspace}

\newcommand{\slevel}{$^2\mathrm{S}_{1/2}$\xspace}
\newcommand{\dlevel}{$^2\mathrm{D}_{5/2}$\xspace}
\newcommand{\groundstate}{$^2\mathrm{S}_{1/2}$\xspace}
\newcommand{\excitedstate}{$^2\mathrm{D}_{5/2}$\xspace}
\newcommand{\clocktrans}{$^2\mathrm{S}_{1/2} \leftrightarrow ^2\mathrm{D}_{5/2}$\xspace}

\newcommand{\ts}[1]{\textsuperscript{#1}}
\newcommand{\eqname}[1]{\tag*{#1}}
\begin{document}
\title{Multi-ion frequency reference using dynamical decoupling}

\author{Lennart Pelzer}
\affiliation{Physikalisch-Technische Bundesanstalt, Bundesallee 100, 38116 Braunschweig, Germany}
\author{Kai Dietze}
\affiliation{Physikalisch-Technische Bundesanstalt, Bundesallee 100, 38116 Braunschweig, Germany}
\affiliation{Institut f^^c3^^bcr Quantenoptik, Leibniz Universit^^c3^^a4t Hannover, Welfengarten 1, 30167 Hannover, Germany}
\author{V^^c3^^adctor Jos^^c3^^a9 Mart^^c3^^adnez-Lahuerta}
\affiliation{Institut f^^c3^^bcr Theoretische Physik, Leibniz Universit^^c3^^a4t Hannover, Appelstra^^c3^^9fe 2, 30167 Hannover }
\affiliation{Institut f^^c3^^bcr Quantenoptik, Leibniz Universit^^c3^^a4t Hannover, Welfengarten 1, 30167 Hannover, Germany}
\author{Ludwig Krinner}
\affiliation{Physikalisch-Technische Bundesanstalt, Bundesallee 100, 38116 Braunschweig, Germany}
\affiliation{Institut f^^c3^^bcr Quantenoptik, Leibniz Universit^^c3^^a4t Hannover, Welfengarten 1, 30167 Hannover, Germany}
\author{Johannes Kramer}
\affiliation{Physikalisch-Technische Bundesanstalt, Bundesallee 100, 38116 Braunschweig, Germany}
\affiliation{Institut f^^c3^^bcr Quantenoptik, Leibniz Universit^^c3^^a4t Hannover, Welfengarten 1, 30167 Hannover, Germany}
\author{Fabian Dawel}
\affiliation{Physikalisch-Technische Bundesanstalt, Bundesallee 100, 38116 Braunschweig, Germany}
\affiliation{Institut f^^c3^^bcr Quantenoptik, Leibniz Universit^^c3^^a4t Hannover, Welfengarten 1, 30167 Hannover, Germany}
\author{Nicolas C. H. Spethmann}
\affiliation{Physikalisch-Technische Bundesanstalt, Bundesallee 100, 38116 Braunschweig, Germany}
    \author{Klemens Hammerer}
\affiliation{Institut f^^c3^^bcr Theoretische Physik, Leibniz Universit^^c3^^a4t Hannover, Appelstra^^c3^^9fe 2, 30167 Hannover }
\author{Piet O. Schmidt  }
\affiliation{Physikalisch-Technische Bundesanstalt, Bundesallee 100, 38116 Braunschweig, Germany}
\affiliation{Institut f^^c3^^bcr Quantenoptik, Leibniz Universit^^c3^^a4t Hannover, Welfengarten 1, 30167 Hannover, Germany}

\date{\today}

\begin{abstract}
We present the experimental realization of a continuous dynamical decoupling scheme which suppresses leading frequency shifts in a multi-ion frequency reference based on \ca. By near-resonant magnetic coupling of the \groundstate\ and \excitedstate\ Zeeman sub-levels using radio-frequency dressing fields, engineered transitions with reduced sensitivity to magnetic-field fluctuations are obtained. A second stage detuned dressing field reduces the influence of amplitude noise in the first stage driving fields and decreases 2\ts{nd}-rank tensor shifts, such as the electric quadrupole shift. Suppression of the quadratic dependence of the quadrupole shift to $\SI{3(2)}{\milli \hertz \per \mu \metre^2}$ and coherence times of \SI{290(20)}{\milli \second} on the optical transition are demonstrated even within a laboratory environment with significant magnetic field noise. Besides removing inhomogeneous line shifts in multi-ion clocks, the demonstrated dynamical decoupling technique may find applications in quantum computing and simulation with trapped ions by a tailored design of decoherence-free subspaces.
\end{abstract}

\maketitle

Decoherence and dephasing of quantum systems through interaction with the environment pose a major challenge in the implementation of quantum sensors, quantum computers and quantum simulators \citep{acin_quantum_2018}. In particular, magnetic and electric fields can cause significant shifts that induce decoherence and dephasing due to their spectral noise properties and inhomogeneity across the quantum system.
For example, in optical clocks based on trapped ions electric field gradients of the trapping potential couple to the electric quadrupole moment of the clock states causing a quadrupole shift (QPS) \citep{itano_external-field_2000}. Several mitigation strategies have been developed to overcome these limitations. Technological solutions in the form of active stabilization and passive shielding have been demonstrated for magnetic fields in a variety of quantum systems \cite{merkel_magnetic_2019, zeng_combined_2023, borkowski_active_2023, duan_simple_2022, altarev_magnetically_2014, dijck_sympathetically_2023, ji_actively_2021, leopold_cryogenic_2019, farolfi_design_2019, devlin_superconducting_2019, wodey_scalable_2020, vovrosh_additive_2018, brandl_cryogenic_2016-1, altarev_large-scale_2015, lee_calculation_2008, mager_magnetic_1970, hanson_magnetically_1965}. In combination with permanent magnets, second long coherence times between magnetically sensitive levels could be observed, e.g. in trapped calcium ions \citep{ruster_long-lived_2016}.
However, most of these technical measures increase size and complexity of a setup, or can not be eliminated by shielding, such as the QPS.
In atomic clocks, mitigation strategies based on atomic properties include choosing magnetic field insensitive Zeeman levels, or averaging over several Zeeman components, which at the same time eliminates the QPS \citep{dube_electric_2005, margolis_hertz-level_2004, chwalla_absolute_2009, dube_evaluation_2013}. The latter can also be eliminated by averaging the transition frequencies of a single transition with three mutually orthogonal magnetic fields applied to the system \citep{itano_external-field_2000, margolis_hertz-level_2004, oskay_measurement_2005, schneider_sub-hertz_2005}, or by applying the magnetic field at an angle of $54.7^\circ$ to the electric field gradient \cite{tan_suppressing_2019, steinel_evaluation_2023}.
However, many of these strategies fail for multi-ion clocks that aim at improving the signal-to-noise ratio of optical ion clocks by employing more than one ion \citep{aharon_robust_2019, keller_controlling_2019, arnold_prospects_2015, herschbach_linear_2012, champenois_ion_2010}, since the shifts are inhomogeneous across the ion crystal, resulting in an effective broadening of the observed clock transition.
A solution is provided by dynamical decoupling (DD), a technique that dates back to Hahn spin echoes in nuclear magnetic resonance spectroscopy \citep{hahn_spin_1950}.

Since then it has been further developed using pulsed \citep{viola_dynamical_1999, valahu_quantum_2022, viola_dynamical_1999} and continuous schemes \citep{valahu_quantum_2022, cai_optimizing_2022, yalcinkaya_continuous_2019, aharon_robust_2019, cohen_continuous_2017, mikelsons_universal_2015, aharon_general_2013, bermudez_robust_2012, cai_robust_2012, chaudhry_decoherence_2012, facchi_unification_2004, viola_dynamical_1998}, and employed experimentally in various systems to eliminate noise sources in quantum simulations \citep{morong_engineering_2023}, quantum computing \citep{barthel_robust_2023, leu_fast_2023-1, fang_crosstalk_2022,pokharel_demonstration_2018, stark_clock_2018, randall_efficient_2015, tan_demonstration_2013, webster_simple_2013, souza_experimental_2012, noguchi_generation_2012, xu_coherence-protected_2012, zalivako_continuous_2023}, quantum memories \citep{wang_single_2021, sinuco-leon_decoherence-free_2021, sinuco-leon_microwave_2019, trypogeorgos_synthetic_2018, laucht_dressed_2017, zhang_protected_2014, golter_protecting_2014, liu_noise-resilient_2013, biercuk_optimized_2009, timoney_quantum_2011, lange_universal_2010}, and quantum sensing \citep{zhang_estimation_2021, baumgart_ultrasensitive_2016, kotler_single-ion_2011, hall_ultrasensitive_2010, pham_enhanced_2012, zhang_estimation_2021, bishof_optical_2013}.
In the context of precision spectroscopy and optical clocks, DD is implemented, e.g. by changing the magnetic field within a Ramsey sequence \citep{lange_coherent_2020} or through precisely timed radio-frequency (rf) pulses that populate different Zeeman states to cancel undesired shifts within each individual interrogation. The latter has been demonstrated for QPS and magnetic field shift in \sr \cite{shaniv_quadrupole_2019} and \lu \cite{kaewuam_hyperfine_2020}, as well as employed to eliminate magnetic field shifts in a precision measurement of local Lorentz violation \citep{yeh_robust_2023, dreissen_improved_2022}. These pulsed schemes can suppress noise sources that vary on time scales slower than the time between pulses.

Here, we demonstrate the suppression of linear Zeeman shifts as well as the QPS on an optical transition in a trapped \ca frequency reference using a cascaded continuous dynamical decoupling (CDD) scheme with up to five ions.  In contrast to similar pulsed schemes \citep{shaniv_quadrupole_2019, kaewuam_hyperfine_2020}, here the levels are continuously coupling with rf-fields to engineer an ensemble of robust artificial transitions by choosing adequate driving parameters \citep{aharon_robust_2019, martinez-lahuerta_quadrupole_2023}. This suppresses inhomogeneous shifts across the ion crystal and mitigates noise up to half the Fourier frequencies corresponding to the rf Rabi frequency coupling of the Zeeman states. These artificial transitions can be treated similar to bare transitions, so e.g., Rabi and Ramsey spectroscopy sequences as well as entangling operations on motional sidebands are possible \citep{martinez-lahuerta_quadrupole_2023}.

\begin{figure*}
\includegraphics[width=0.95\linewidth]{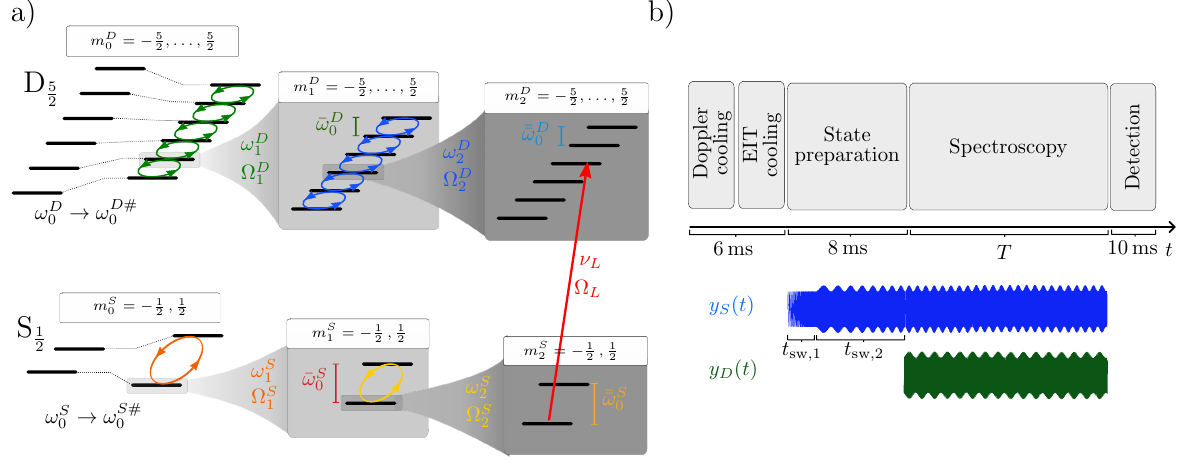}
\caption{a) Continuous dynamical decoupling scheme. Near resonant rf-coupling with driving frequency $\omega_j^i$ and strength $\Omega_j^i$ results in an ensemble of $2(J_j^i+1)$ dressed states with spacing $\bar{\omega}_j^i$ and numbering $m_j^i$ where $j=1,2$ marks the 1\ts{st} or 2\ts{nd} layer of dressing and $i=S,D$ the Zeeman manifold. Here, the quantum numbers $J_1^S =1/2 = J_2^S$ and $J_1^D =5/2 = J_2^D$ match these of the spin of the underlying bare atomic states. Cross-coupling and  Bloch-Siegert shifts result in a shift of the bare states' transition frequency $\omega_0^j \rightarrow \omega_0^{j\#}$. The doubly-dressed transitions are probed with $\SI{729}{\nano\metre}$ light at frequency $\nu_L$ and coupling strength $\Omega_L$. Adapted from \cite{pelzer_robust_2023} b) Sketch of the spectroscopy sequence with typical durations. The lower part shows a sketch of the employed CDD waveforms with signal amplitudes $y_i$. Two successive frequency sweeps transfer the population adiabatically from the \slevel\ via the first dressed to the doubly-dressed target state. \label{fig:CDD_scheme}}
\end{figure*}

In the following we briefly summarize the CDD technique employed here \citep{aharon_robust_2019, martinez-lahuerta_quadrupole_2023} and which are shown schematically in Fig~\ref{fig:CDD_scheme}.
The Zeeman manifolds of ground \slevel\ and excited \dlevel\ levels are energetically split by a static external magnetic field with magnitude $B_0$ in $z$-direction
\begin{equation}
H_0 = \sum_{i=S,D} g_i \mu_B B_0 J_z^i = \sum_{i=S,D}\omega_0^i J_z^i
\end{equation}
Here, the gyromagnetic factor $g_i$ and spin operator $z$-component $J_z^i$ for each ensemble $i \in [\text{S}_\frac{1}{2}, \text{D}_\frac{5}{2}]$ as well as the Bohr magneton $\mu_B$ are used.
These so-called bare atomic levels are coupled by near-resonant rf-fields of the form
\begin{multline}\label{eq:drive}
H_{\text{rf}} = \sum_{i=S,D} g_i \Omega_1^i \cos(\omega_1^it)J_x^i \\
-g_i\Omega_2^i\sin(\omega_1^i t) \cos(\omega_2^i t) J_x^i \; .
\end{multline}
The first drive (first line in Eq.~\ref{eq:drive})) with frequency $\omega_1^i = \omega_0^i-\Delta_1^i$ and amplitude $\Omega_1^i$ produces a ladder of dressed states with frequency splitting
\begin{equation}
\bar{\omega}_0^i =\sqrt{(g_i\Omega_1^i/2)^2  + (\Delta_1^i)^2} \; .
\end{equation}
The second stage drive (2\ts{nd} line in Eq.~\ref{eq:drive})  with frequency $\omega_2^i = \bar{\omega}_0^i-\Delta_2^i$ and amplitude $\Omega_2^i$ couples the first stage dressed states analogously to the first drive if the hierarchy of coupling frequencies $\omega_2^i \ll \omega_1^i$ is fulfilled. Here, colinear alignment of all rf-drives is assumed for simplicity. For a detailed derivation of the general geometry we refer the reader to \cite{martinez-lahuerta_quadrupole_2023}. The two-stage design serves two purposes. Firstly, temporal or spatial drive strength variations $\delta \Omega_1^i(\vec{r},t)$ will result in undesired energy shifts of the resulting states. Choosing a weaker second drive can mitigate these shifts, since assuming a constant fractional variation results in smaller absolute frequency shifts. Secondly, the additional drive adds control parameters to suppress Zeeman and 2\ts{nd}-rank tensor shifts simultaneously as will be shown below.
The so-called mixing angle $\cos( \theta_j^i) = \Delta_j^i/\bar{\omega}_j^i$ with $j\in[1,2]$ and $ i\in[S_{\frac{1}{2}},D_{\frac{5}{2}}]$ determines the suppression of these shifts for each stage and manifold. In a doubly-dressed basis, the response of the system to sufficiently slow magnetic field variations $\delta B(\vec{r},t)$ can be described by
\begin{equation}\label{eq:B_supress}
V_{\delta B} = \sum_{i=S,D} \cos(\theta_1^i) \cos(\theta_2^i) g_i \mu_B \delta B(\vec{r},t)J_z^i
\end{equation}
and the quadrupole shift of the excited level by \citep{martinez-lahuerta_quadrupole_2023}
\begin{multline}\label{eq:QPS_supress}
V_{Q} =  \left(1-3\cos^2(\theta_1^D) \right) \left(1-3\cos^2(\theta_2^D) \right) \\
\times \Theta\frac{3}{8}\frac{\partial E_z}{\partial z} \left( J^D(J^D+1) -3(J_z^D)^2 \right)\, ,
\end{multline}
using the spin of the D-level $J^D$. The magnitude of the shift depends on the quadrupole moment $\Theta$ and the electric field gradient $\partial E/\partial z$ \citep{itano_external-field_2000}, which is given by the joint potential of the trap and neighbouring ions. It is worth noting that the suppression of both shifts is independent for both stages. This allows e.g. suppression of the Zeeman shift using a resonant drive with $\theta_1^S=\theta_2^S=\theta_1^D=0$ for all drives except the 2\ts{nd}-stage $D$-level drive. QPS suppression can be attained using the so-called magic mixing angle on this last drive with $\cos^2(\theta_2^D)=\frac{1}{3}$ at the expense of a reduced Zeeman shift suppression.
The achievable degree of suppression depends on the deviation of the angles $\theta_j^i$ from their ideal values. The detunings $\Delta_j^i$ are affected by magnetic field fluctuations acting on the bare atomic states, the QPS of the D states, frequency shifts from off-resonant cross-coupling of the S drive on the D level and vice versa, as well as the correction of counter-rotating terms in the rotating wave approximation (Bloch-Siegert shift) \cite{aharon_robust_2019, martinez-lahuerta_quadrupole_2023}.
Since $\theta_j^i$ depends inversely proportional on $\Omega_j^i$, the scheme becomes more robust against fluctuations in magnitude and noise frequency of these shifts for higher coupling strength at the expense of larger frequency shifts caused by amplitude noise of the rf-drives, assuming a fixed fractional stability. The parameter set for optimum suppression is thus a trade-off between detuning fluctuations and driving strength, requiring a detailed knowledge of the noise properties of both.

In the setup presented here, the magnetic field noise causes line broadening on the order of $\nu_{\delta B} \lesssim \SI{1}{\kilo \hertz}$ with a timescale on the order of milliseconds given by the mains supply and its harmonics. In contrast, the QPS is near static with an inhomogeneous shift of $\nu_{\text{QPS}}\lesssim \SI{100}{\hertz}$ for typical trapping potentials.
Assuming a realistic mixing angle mismatch of $\delta \theta = 1\%$, a Zeeman shift suppression on the order of $\cos(\theta_1^i) \cos(\theta_2^i)  \approx 10^{-4}$ (compare Eq.~\ref{eq:B_supress}) for a doubly resonant drive and $\left(1-3\cos^2(\theta_2^D) \right) \approx 3\times 10^{-2}$ QPS suppression (Eq.~\ref{eq:QPS_supress}) can be expected for small deviations. The sensitivities of the dressed states to magnetic field changes can be fine-tuned separately for S and D-levels, such that a residual linear sensitivity due to the magic-angle detuned second D-drive can be partially compensated with a detuned second S-drive.

The majority of the experimental setup is described in detail elswhere \cite{hannig_towards_2019, pelzer_robust_2023}, therefore only key components are summarized here. \ca ions are confined in a segmented Paul trap \cite{pyka_high-precision_2014} with low excess micromotion even for axially extended crystals \cite{hannig_towards_2019, keller_controlling_2019} as well as low motional heating due to on-chip filtering and a low noise voltage supply \citep{beev_low-drift_2017}. All lasers needed for cooling, detection and state preparation are locked to a wavemeter\footnote{High Finesse WS U10} and most beams are supplied via fibre-coupled components to obtain a robust setup with small footprint. The individual beams are switched and frequency steered by acousto-optic modulators (AOM) that are controlled by a pulse sequencer \cite{hannig_towards_2019}. Light shifts during probing of the clock transition are minimized by using shutters in all beam paths.

The interrogation laser for the \clocktrans\ transition at \SI{729}{nm} is based on an amplified extended cavity diode laser\footnote{TA pro Toptica}, which is first pre-stabilized to a \SI{10}{cm} cavity resulting in a fractional linewidth of \SI{170(4)}{\hertz}. Further reduction of the laser linewidth is established by transfer locking \cite{stenger_ultraprecise_2002, scharnhorst_high-bandwidth_2015} to a highly stable reference cavity \cite{matei_1.5_2017} via a fibre-based frequency comb. An independent measurement revealed an upper limit for the fractional frequency instability of the \ca clock laser against the reference laser of $\sigma_y < 2\times10^{-15}$ for $\tau> \qty{1}{\second}$. The reference laser reaches typically an instability of  $5\times10^{-17}$ for up to  \SI{10}{\second} and a drift of \SI{50}{\micro\hertz \per \second}. Therefore, the employed laser is not limiting for the following measurements.

A single aspheric lens is used for collecting fluorescence light of the ions. It is focused simultaneously onto a photomultiplier tube (PMT) and a scientific complementary metal-oxide-semiconductor (sCMOS) camera using a beam splitter. The PMT allows for discrimination of bright and dark states within \SI{100}{\micro\second} detection time \cite{hannig_towards_2019}. Using the camera, spatially resolved detection of single ion excitation in a linear ion crystal is possible at the cost of a longer detection time of \SI{10}{\milli\second}. For this, the counts in a region of interest around each ion are analysed.

Three pairs of orthogonal magnetic field coils generate a static offset field of \qty{357}{\micro \tesla} along the axial trap direction resulting in \qty{10}{\mega \hertz} splitting of the \slevel\ levels and an average quadratic Zeeman shift of \SI{1.37}{\hertz} \citep{chwalla_absolute_2009}. The axial magnetic field component is actively stabilized against slow drifts using a fluxgate sensor and feedback to the current of the coils. Mains power synchronous magnetic field noise limits the linewidth of the \clocktrans\ clock transition typically to $\Delta \nu \approx \SI{100}{\hertz}$. In addition, on  timescales of ten seconds fluctuations on the order of $\vert \Delta  B_0 \vert \approx \SI{5}{\nano \tesla}$ are observed.

Resonant magnetic field coils are employed for the CDD to shape and enhance the generated coupling waveforms. These rf-coils consist of two separate LCR circuits with tunable capacitors to match the resonance frequency of the Zeeman ensembles. The quality factors $Q_S = 8.59(3)$, $Q_D =15.95(4)$ of the coils are chosen as a compromise between required field strength and minimizing signal distortion by its transfer function. Duty cycle and laboratory temperature change the temperature-dependent transfer function of the coils, resulting in a variation in coupling strength. Therefore, the capacitive and resistive parts are mounted on a temperature-controlled baseplate and the coils are designed to minimize warm-up during operation.

The mounts are placed on translation stages and positioned in a distance of approximately \SI{5}{\centi\meter} to the ion crystal in an inverted viewport outside the vacuum chamber. The signal for each coil is supplied via an inductively-coupled, impedance-matched in-coupling coil, which is driven by an amplifier. The signal source is a two-channel arbitrary voltage generator\footnote{Keysight 33622A} (AWG) allowing flexible waveforms, e.g. for state preparation using a rapid adiabatic passage (see below). To avoid distortion of the two-tone signal by the phase and amplitude response of the coils, their transfer function is measured and pre-compensated in the AWG waveform by applying the inverse transfer function to the waveform (compare Eq.~\ref{eq:transfer_function}).

\begin{figure}[t]
\includegraphics[width=0.95\linewidth]{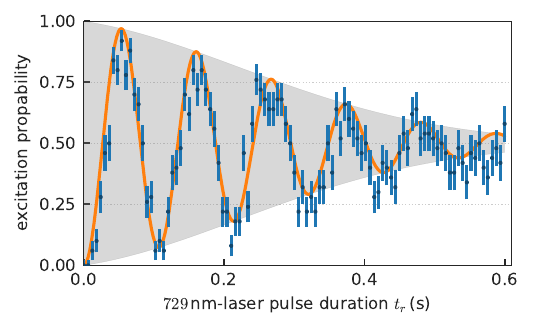}
\caption{\SI{729}{\nano \metre} pulse time scan of a doubly-dressed artificial transition. The strong Zeeman shift suppression with resonant driving parameters allows a long coherence time. The data is in good agreement with a combined Gaussian and exponential decay model $P_e(t_r)= \frac{1}{2} \left( 1-\cos( 2\pi \Omega_L t_r ) e^{-t_r/\Gamma}e^{-t_r^2/(2\gamma^2)}\right)$ with given natural decay $\Gamma= \SI{1.168}{\second}$ \citep{kreuter_experimental_2005} and $\gamma= \SI{0.29(2)}{\second}$. The Gaussian decay shape indicates long noise correlation times with respect to the duration of an individual data point \citep{monz_quantum_2011}. \label{flop}}
\label{fig:CDDflop}
\end{figure}
The CDD sequence begins with the preparation of the target state in the dressed S-levels. This is achieved through an adiabatic ramp from the bare ground-state to the dressed level by applying a tailored waveform to the S-coils (see supplementary material Eqs.~\ref{eq:first_sweep} - \ref{eq:second_stage}). A continuous transition without phase jumps from the preparation to the final second-stage waveform is crucial to maximize the contrast. The D-level waveform can be switched on instantaneously, since the D state is not populated and the cross-driving effect of the D-level drive onto the population in the S-level is negligible. Spectroscopy within the CDD spectrum with the \slevel\ to \dlevel\ laser at \SI{729}{\nano \metre} is performed after the applied rf-drives have reached their steady-state values. As long as the optical coupling is weak compared to the rf-coupling $\Omega_L \ll \Omega_i$ the fast dynamics of the manifold averages out and the artificial transitions can be treated analogously to a natural \ca clock transition. In the following the $\ket{S_{1/2}, m_0^S=-\frac{1}{2}, m_1^S=\frac{1}{2}, m_2^S= \frac{1}{2}} \leftrightarrow \ket{D_{5/2},m_0^D=-\frac{3}{2},m_1^D= \frac{1}{2}, m_2^D=\frac{1}{2}}$
transition (compare Fig.~\ref{fig:CDD_scheme}) is probed with the clock laser directed along the static magnetic field and trap axis. With the parameters given in Table~\ref{tab:CDD_parameter}, a magnetic field shift sensitivity of $\SI{-1.2(1)e-4}{\hertz \per \nano \tesla^2}$ was determined in a separate measurement. This represents an almost four orders of magnitude improvement over the sensitivity of the bare states of \SI{5.6}{\hertz \per \nano \tesla} for typical field variations of \SI{5}{\nano \tesla} and is comparable to clock species with magnetic-field sensitivity only from nuclear spin, such as \al and \indium \cite{rosenband_observation_2007, peik_laser_1994}. Therefore, a coherence time of \SI{0.29(2)}{\second} can be observed on one of the artificial transitions (see Fig.~\ref{fig:CDDflop}). The optical coupling strength of each transition between dressed states depends on the selection rules of the underlying Zeeman transition as well as on mixing angles and quantum numbers of the dressed state \cite{martinez-lahuerta_quadrupole_2023}.

\begin{figure}[t]
\includegraphics[scale=1.0]{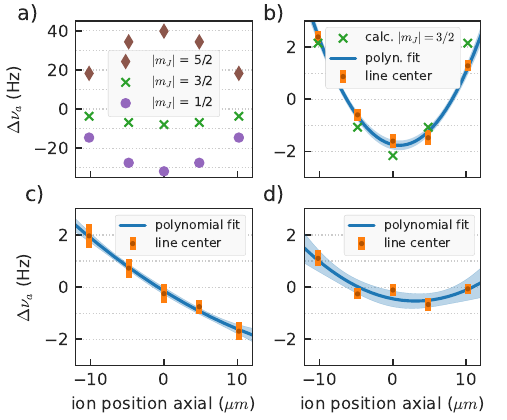}
\caption{Comparison of the relative frequency shifts of the artificial clock transition along a linear five-ion crystal for different CDD-parameters. The individual line centers are obtained from a scan of the clock laser with fitting of individual peaks using ion-resolved camera data. a) Calculated QPS for five-ion crystal for bare Zeeman levels. Ion positions are simulated in a pure harmonic potential with given trapping frequencies to extract field gradients at the ion positions. b) Doubly resonant parameter set ($\Delta_j^i =0 \quad \forall i,j$). Tensorial shifts are of equal size compared to the calculated $\vert m_J \vert = \frac{3}{2}$ bare level shift (green crosses). c) Magic angle detuned second stage D-level CDD drive. The visible linear dependency arises from magnetic field inhomogeneity in combination with lower Zeeman shift suppression compared to (b).  d) Same CDD drive as for (c), but optimized offset B-field for field inhomogeneity compensation.  \label{fig:QPS}}
\end{figure}

In order to demonstrate the capabilities of this scheme we probe the line shifts of a linear five-ion crystal. Such a system is a stringent test for a multi-ion frequency reference. In Fig.~\ref{fig:QPS} the comparison of inhomogeneous line shifts along the crystal for two different CDD-parameter sets is shown. The frequencies of the first stage rf-fields are held close to the respective splitting of the Zeeman states $\Delta_1^S = 0= \Delta_1^D$. The detuning of the 2\ts{nd} D-level stage is set for the two cases $\Delta_2^D=0$ and $\Delta_2^D=1/\sqrt{8}g_D \Omega_2^D$, corresponding to the resonant and magic mixing angle, respectively. In the first case only Zeeman shifts are suppressed, the second also suppresses 2\ts{nd}-rank tensor shifts at the cost of reduced Zeeman shift suppression. A relative QPS suppression factor of 12 was measured between resonant and magic detuned parameter set (Fig.~\ref{fig:QPS} b, c), measured by the quadratic fitting parameter of $\SI{3(2)}{\milli \hertz \per \mu \metre^2}$. Simultaneously, the linear term increases by a factor of 3, caused by lower suppression of the magnetic field inhomogeneity. In a third setting, the CDD parameters are fine-tuned by an offset of the static B-field such that the imperfections in residual inhomogeneous static and dynamic B-field partially compensate, which allows to suppress the overall frequency spread across the $\SI{20}{\micro \meter}$ long crystal to \SI{1.8(7)}{\hertz} (Fig.~\ref{fig:QPS} d).
Further optimization by variation of all CDD parameters is likely to improve the results. However, technical measures for enhanced homogeneity and stability of the fields will loosen the requirements to find the optimal parameters.
\begin{figure}
\includegraphics[width=0.95\linewidth]{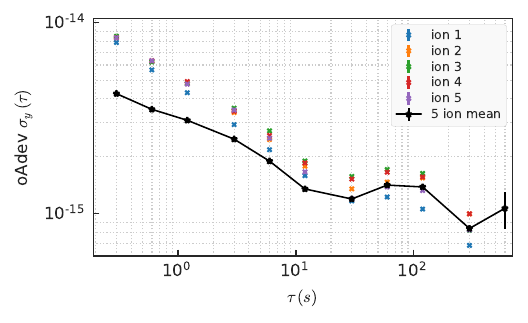}
\caption{Overlapping Allan deviation of the relative frequency stability of a linear five-ion crystal. Two-point sampling Rabi interrogation with $\SI{100}{\milli \second}$ interrogation time and duty cycle of $33\%$ was used. The excitation probability of the individual ions was measured using an sCMOS camera. The sensitivity to small dynamic or offset magnetic field variations depends on the absolute field value. Therefore, inhomogeneous fields cause stability differences for the individual ions.  \label{fig:allan_5_ions}}
\end{figure}
The slow temporal variations of the leading frequency shifts, namely residual amplitude fluctuations of the CDD-drive and of the magnetic field, are measured to be on the level of $\vert \frac{\delta \Omega_i}{\Omega_i} \vert <6 \times 10^{-5}$ and $\vert\frac{\delta B_0 }{ B_0 }\vert <2 \times 10^{-5}$, respectively. These were determined by observing the splitting of bare and first-stage dressed states over several \SI{100}{\second}.

The frequency instability of the doubly-dressed transition of a five-ion crystal is displayed in Fig.~\ref{fig:allan_5_ions}. With the chosen parameters, a gain in statistical uncertainty due to increased ion number is given up to \SI{10}{\second} averaging time before residual magnetic field drift compromise the instability. The Zeeman shift sensitivity depends on the absolute field at the position of each ion, therefore the stability varies for the individual ions. However, the short-term instability of $3\times 10^{-15}/\sqrt{\tau}$ is comparable to \ca clocks with multi-layer magnetic shielding \cite{huang_nearly_2021}. Increasing the interrogation duty cycle from $T/T_c = 33\%$ to $ \approx 90\%$ will be possible with a faster experimental control system together with faster camera detection and readout.

In conclusion, we have demonstrated the suppression of linear Zeeman shifts and QPS on an artificial clock transition through CDD. The presented CDD scheme is especially useful for transitions with strong electronic linear Zeeman shift. This lifts the need for elaborate shielding and reduces the statistical uncertainty in clocks by enabling longer interrogation times. By reducing magnetic field noise and drive field inhomogeneity, as well as implementing faster camera detection, a larger number of ions would enable this frequency reference to be used as a flywheel in composite, multi-species clocks \cite{hume_probing_2016, dorscher_dynamical_2020, rosenband_exponential_2013, borregaard_efficient_2013}.
To the best of our knowledge, this is the first time a 2-stage CDD scheme has been successfully implemented, which might prove useful for other quantum sensors or qubit systems that are limited by magnetic field sensitivity.

\section{acknowledgments}
We thank PTB^^e2^^80^^99s unit-of-length working group for
providing the stable silicon referenced laser source.
Fruitful discussions with Nati Aharon, Alex Retzker
and the group of Roee Ozeri are acknowledged. This joint research
project was financially supported by the State of
Lower Saxony, Hanover, Germany through Nieders^^c3^^a4chsisches
Vorab and by the Deutsche Forschungsgemeinschaft
(DFG, German Research Foundation)
^^e2^^80^^93 Project-ID 274200144 ^^e2^^80^^93 SFB 1227 (DQ-mat,
projects A06 and B03). This project also received
funding from the European Metrology Programme
for Innovation and Research (EMPIR) cofinanced
by the Participating 5 States and from the European
Union^^e2^^80^^99s Horizon 2020 research and innovation
programme (Project No. 20FUN01 TSCAC).

\bibliography{main.bib}

\begin{thebibliography}{99}%
\makeatletter
\providecommand \@ifxundefined [1]{%
 \@ifx{#1\undefined}
}%
\providecommand \@ifnum [1]{%
 \ifnum #1\expandafter \@firstoftwo
 \else \expandafter \@secondoftwo
 \fi
}%
\providecommand \@ifx [1]{%
 \ifx #1\expandafter \@firstoftwo
 \else \expandafter \@secondoftwo
 \fi
}%
\providecommand \natexlab [1]{#1}%
\providecommand \enquote  [1]{``#1''}%
\providecommand \bibnamefont  [1]{#1}%
\providecommand \bibfnamefont [1]{#1}%
\providecommand \citenamefont [1]{#1}%
\providecommand \href@noop [0]{\@secondoftwo}%
\providecommand \href [0]{\begingroup \@sanitize@url \@href}%
\providecommand \@href[1]{\@@startlink{#1}\@@href}%
\providecommand \@@href[1]{\endgroup#1\@@endlink}%
\providecommand \@sanitize@url [0]{\catcode `\\12\catcode `\$12\catcode
  `\&12\catcode `\#12\catcode `\^12\catcode `\_12\catcode `\%12\relax}%
\providecommand \@@startlink[1]{}%
\providecommand \@@endlink[0]{}%
\providecommand \url  [0]{\begingroup\@sanitize@url \@url }%
\providecommand \@url [1]{\endgroup\@href {#1}{\urlprefix }}%
\providecommand \urlprefix  [0]{URL }%
\providecommand \Eprint [0]{\href }%
\providecommand \doibase [0]{https://doi.org/}%
\providecommand \selectlanguage [0]{\@gobble}%
\providecommand \bibinfo  [0]{\@secondoftwo}%
\providecommand \bibfield  [0]{\@secondoftwo}%
\providecommand \translation [1]{[#1]}%
\providecommand \BibitemOpen [0]{}%
\providecommand \bibitemStop [0]{}%
\providecommand \bibitemNoStop [0]{.\EOS\space}%
\providecommand \EOS [0]{\spacefactor3000\relax}%
\providecommand \BibitemShut  [1]{\csname bibitem#1\endcsname}%
\let\auto@bib@innerbib\@empty
%</preamble>
\bibitem [{\citenamefont {Ac{\'i}n}\ \emph {et~al.}(2018)\citenamefont
  {Ac{\'i}n}, \citenamefont {Bloch}, \citenamefont {Buhrman}, \citenamefont
  {Calarco}, \citenamefont {Eichler}, \citenamefont {Eisert}, \citenamefont
  {{Daniel Esteve}}, \citenamefont {Gisin}, \citenamefont {Glaser},
  \citenamefont {Jelezko}, \citenamefont {Kuhr}, \citenamefont {Lewenstein},
  \citenamefont {Riedel}, \citenamefont {Schmidt}, \citenamefont {Thew},
  \citenamefont {Wallraff}, \citenamefont {Walmsley},\ and\ \citenamefont
  {Wilhelm}}]{acin_quantum_2018}%
  \BibitemOpen
  \bibfield  {author} {\bibinfo {author} {\bibfnamefont {A.}~\bibnamefont
  {Ac{\'i}n}}, \bibinfo {author} {\bibfnamefont {I.}~\bibnamefont {Bloch}},
  \bibinfo {author} {\bibfnamefont {H.}~\bibnamefont {Buhrman}}, \bibinfo
  {author} {\bibfnamefont {T.}~\bibnamefont {Calarco}}, \bibinfo {author}
  {\bibfnamefont {C.}~\bibnamefont {Eichler}}, \bibinfo {author} {\bibfnamefont
  {J.}~\bibnamefont {Eisert}}, \bibinfo {author} {\bibnamefont {{Daniel
  Esteve}}}, \bibinfo {author} {\bibfnamefont {N.}~\bibnamefont {Gisin}},
  \bibinfo {author} {\bibfnamefont {S.~J.}\ \bibnamefont {Glaser}}, \bibinfo
  {author} {\bibfnamefont {F.}~\bibnamefont {Jelezko}}, \bibinfo {author}
  {\bibfnamefont {S.}~\bibnamefont {Kuhr}}, \bibinfo {author} {\bibfnamefont
  {M.}~\bibnamefont {Lewenstein}}, \bibinfo {author} {\bibfnamefont {M.~F.}\
  \bibnamefont {Riedel}}, \bibinfo {author} {\bibfnamefont {P.~O.}\
  \bibnamefont {Schmidt}}, \bibinfo {author} {\bibfnamefont {R.}~\bibnamefont
  {Thew}}, \bibinfo {author} {\bibfnamefont {A.}~\bibnamefont {Wallraff}},
  \bibinfo {author} {\bibfnamefont {I.}~\bibnamefont {Walmsley}},\ and\
  \bibinfo {author} {\bibfnamefont {F.~K.}\ \bibnamefont {Wilhelm}},\ }\href
  {https://doi.org/10.1088/1367-2630/aad1ea} {\bibfield  {journal} {\bibinfo
  {journal} {New J. Phys.}\ }\textbf {\bibinfo {volume} {20}},\ \bibinfo
  {pages} {080201} (\bibinfo {year} {2018})}\BibitemShut {NoStop}%
\bibitem [{\citenamefont {Itano}(2000)}]{itano_external-field_2000}%
  \BibitemOpen
  \bibfield  {author} {\bibinfo {author} {\bibfnamefont {W.~M.}\ \bibnamefont
  {Itano}},\ }\href {https://doi.org/10.6028/jres.105.065} {\bibfield
  {journal} {\bibinfo  {journal} {J. Res. NIST}\ }\textbf {\bibinfo {volume}
  {105}},\ \bibinfo {pages} {829} (\bibinfo {year} {2000})}\BibitemShut
  {NoStop}%
\bibitem [{\citenamefont {Merkel}\ \emph {et~al.}(2019)\citenamefont {Merkel},
  \citenamefont {Thirumalai}, \citenamefont {Tarlton}, \citenamefont
  {Sch{\"a}fer}, \citenamefont {Ballance}, \citenamefont {Harty},\ and\
  \citenamefont {Lucas}}]{merkel_magnetic_2019}%
  \BibitemOpen
  \bibfield  {author} {\bibinfo {author} {\bibfnamefont {B.}~\bibnamefont
  {Merkel}}, \bibinfo {author} {\bibfnamefont {K.}~\bibnamefont {Thirumalai}},
  \bibinfo {author} {\bibfnamefont {J.~E.}\ \bibnamefont {Tarlton}}, \bibinfo
  {author} {\bibfnamefont {V.~M.}\ \bibnamefont {Sch{\"a}fer}}, \bibinfo
  {author} {\bibfnamefont {C.~J.}\ \bibnamefont {Ballance}}, \bibinfo {author}
  {\bibfnamefont {T.~P.}\ \bibnamefont {Harty}},\ and\ \bibinfo {author}
  {\bibfnamefont {D.~M.}\ \bibnamefont {Lucas}},\ }\href
  {https://doi.org/10.1063/1.5080093} {\bibfield  {journal} {\bibinfo
  {journal} {Rev. Sci. Instrum.}\ }\textbf {\bibinfo {volume} {90}},\ \bibinfo
  {pages} {044702} (\bibinfo {year} {2019})}\BibitemShut {NoStop}%
\bibitem [{\citenamefont {Zeng}\ \emph {et~al.}(2023)\citenamefont {Zeng},
  \citenamefont {Ma}, \citenamefont {Hu}, \citenamefont {Zhang}, \citenamefont
  {Hao}, \citenamefont {Zhang}, \citenamefont {Huang}, \citenamefont {Guan},\
  and\ \citenamefont {Gao}}]{zeng_combined_2023}%
  \BibitemOpen
  \bibfield  {author} {\bibinfo {author} {\bibfnamefont {M.}~\bibnamefont
  {Zeng}}, \bibinfo {author} {\bibfnamefont {Z.}~\bibnamefont {Ma}}, \bibinfo
  {author} {\bibfnamefont {R.}~\bibnamefont {Hu}}, \bibinfo {author}
  {\bibfnamefont {B.}~\bibnamefont {Zhang}}, \bibinfo {author} {\bibfnamefont
  {Y.}~\bibnamefont {Hao}}, \bibinfo {author} {\bibfnamefont {H.}~\bibnamefont
  {Zhang}}, \bibinfo {author} {\bibfnamefont {Y.}~\bibnamefont {Huang}},
  \bibinfo {author} {\bibfnamefont {H.}~\bibnamefont {Guan}},\ and\ \bibinfo
  {author} {\bibfnamefont {K.}~\bibnamefont {Gao}},\ }\bibfield  {journal}
  {\bibinfo  {journal} {Chinese Phys. B}\ }\href
  {https://doi.org/10.1088/1674-1056/acf5d5} {10.1088/1674-1056/acf5d5}
  (\bibinfo {year} {2023})\BibitemShut {NoStop}%
\bibitem [{\citenamefont {Borkowski}\ \emph {et~al.}(2023)\citenamefont
  {Borkowski}, \citenamefont {Reichs{\"o}llner}, \citenamefont {Thekkeppatt},
  \citenamefont {Barb{\'e}}, \citenamefont {{van Roon}}, \citenamefont {{van
  Druten}},\ and\ \citenamefont {Schreck}}]{borkowski_active_2023}%
  \BibitemOpen
  \bibfield  {author} {\bibinfo {author} {\bibfnamefont {M.}~\bibnamefont
  {Borkowski}}, \bibinfo {author} {\bibfnamefont {L.}~\bibnamefont
  {Reichs{\"o}llner}}, \bibinfo {author} {\bibfnamefont {P.}~\bibnamefont
  {Thekkeppatt}}, \bibinfo {author} {\bibfnamefont {V.}~\bibnamefont
  {Barb{\'e}}}, \bibinfo {author} {\bibfnamefont {T.}~\bibnamefont {{van
  Roon}}}, \bibinfo {author} {\bibfnamefont {K.}~\bibnamefont {{van Druten}}},\
  and\ \bibinfo {author} {\bibfnamefont {F.}~\bibnamefont {Schreck}},\ }\href
  {https://doi.org/10.1063/5.0143825} {\bibfield  {journal} {\bibinfo
  {journal} {Rev. Sci. Instrum.}\ }\textbf {\bibinfo {volume} {94}},\ \bibinfo
  {pages} {073202} (\bibinfo {year} {2023})}\BibitemShut {NoStop}%
\bibitem [{\citenamefont {Duan}\ \emph {et~al.}(2022)\citenamefont {Duan},
  \citenamefont {Wu}, \citenamefont {Lin},\ and\ \citenamefont
  {Yang}}]{duan_simple_2022}%
  \BibitemOpen
  \bibfield  {author} {\bibinfo {author} {\bibfnamefont {Z.-X.}\ \bibnamefont
  {Duan}}, \bibinfo {author} {\bibfnamefont {W.-T.}\ \bibnamefont {Wu}},
  \bibinfo {author} {\bibfnamefont {Y.-T.}\ \bibnamefont {Lin}},\ and\ \bibinfo
  {author} {\bibfnamefont {S.-J.}\ \bibnamefont {Yang}},\ }\href
  {https://doi.org/10.1063/5.0119778} {\bibfield  {journal} {\bibinfo
  {journal} {Rev. Sci. Instrum.}\ }\textbf {\bibinfo {volume} {93}},\ \bibinfo
  {pages} {123201} (\bibinfo {year} {2022})}\BibitemShut {NoStop}%
\bibitem [{\citenamefont {Altarev}\ \emph {et~al.}(2014)\citenamefont
  {Altarev}, \citenamefont {Babcock}, \citenamefont {Beck}, \citenamefont
  {Burghoff}, \citenamefont {Chesnevskaya}, \citenamefont {Chupp},
  \citenamefont {Degenkolb}, \citenamefont {Fan}, \citenamefont {Fierlinger},
  \citenamefont {Frei}, \citenamefont {Gutsmiedl}, \citenamefont
  {{Knappe-Gr{\"u}neberg}}, \citenamefont {Kuchler}, \citenamefont {Lauer},
  \citenamefont {Link}, \citenamefont {Lins}, \citenamefont {Marino},
  \citenamefont {McAndrew}, \citenamefont {Niessen}, \citenamefont {Paul},
  \citenamefont {Petzoldt}, \citenamefont {Schl{\"a}pfer}, \citenamefont
  {Schnabel}, \citenamefont {Sharma}, \citenamefont {Singh}, \citenamefont
  {Stoepler}, \citenamefont {Stuiber}, \citenamefont {Sturm}, \citenamefont
  {Taubenheim}, \citenamefont {Trahms}, \citenamefont {Voigt},\ and\
  \citenamefont {Zechlau}}]{altarev_magnetically_2014}%
  \BibitemOpen
  \bibfield  {author} {\bibinfo {author} {\bibfnamefont {I.}~\bibnamefont
  {Altarev}}, \bibinfo {author} {\bibfnamefont {E.}~\bibnamefont {Babcock}},
  \bibinfo {author} {\bibfnamefont {D.}~\bibnamefont {Beck}}, \bibinfo {author}
  {\bibfnamefont {M.}~\bibnamefont {Burghoff}}, \bibinfo {author}
  {\bibfnamefont {S.}~\bibnamefont {Chesnevskaya}}, \bibinfo {author}
  {\bibfnamefont {T.}~\bibnamefont {Chupp}}, \bibinfo {author} {\bibfnamefont
  {S.}~\bibnamefont {Degenkolb}}, \bibinfo {author} {\bibfnamefont
  {I.}~\bibnamefont {Fan}}, \bibinfo {author} {\bibfnamefont {P.}~\bibnamefont
  {Fierlinger}}, \bibinfo {author} {\bibfnamefont {A.}~\bibnamefont {Frei}},
  \bibinfo {author} {\bibfnamefont {E.}~\bibnamefont {Gutsmiedl}}, \bibinfo
  {author} {\bibfnamefont {S.}~\bibnamefont {{Knappe-Gr{\"u}neberg}}}, \bibinfo
  {author} {\bibfnamefont {F.}~\bibnamefont {Kuchler}}, \bibinfo {author}
  {\bibfnamefont {T.}~\bibnamefont {Lauer}}, \bibinfo {author} {\bibfnamefont
  {P.}~\bibnamefont {Link}}, \bibinfo {author} {\bibfnamefont {T.}~\bibnamefont
  {Lins}}, \bibinfo {author} {\bibfnamefont {M.}~\bibnamefont {Marino}},
  \bibinfo {author} {\bibfnamefont {J.}~\bibnamefont {McAndrew}}, \bibinfo
  {author} {\bibfnamefont {B.}~\bibnamefont {Niessen}}, \bibinfo {author}
  {\bibfnamefont {S.}~\bibnamefont {Paul}}, \bibinfo {author} {\bibfnamefont
  {G.}~\bibnamefont {Petzoldt}}, \bibinfo {author} {\bibfnamefont
  {U.}~\bibnamefont {Schl{\"a}pfer}}, \bibinfo {author} {\bibfnamefont
  {A.}~\bibnamefont {Schnabel}}, \bibinfo {author} {\bibfnamefont
  {S.}~\bibnamefont {Sharma}}, \bibinfo {author} {\bibfnamefont
  {J.}~\bibnamefont {Singh}}, \bibinfo {author} {\bibfnamefont
  {R.}~\bibnamefont {Stoepler}}, \bibinfo {author} {\bibfnamefont
  {S.}~\bibnamefont {Stuiber}}, \bibinfo {author} {\bibfnamefont
  {M.}~\bibnamefont {Sturm}}, \bibinfo {author} {\bibfnamefont
  {B.}~\bibnamefont {Taubenheim}}, \bibinfo {author} {\bibfnamefont
  {L.}~\bibnamefont {Trahms}}, \bibinfo {author} {\bibfnamefont
  {J.}~\bibnamefont {Voigt}},\ and\ \bibinfo {author} {\bibfnamefont
  {T.}~\bibnamefont {Zechlau}},\ }\href {https://doi.org/10.1063/1.4886146}
  {\bibfield  {journal} {\bibinfo  {journal} {Rev. Sci. Instrum.}\ }\textbf
  {\bibinfo {volume} {85}},\ \bibinfo {pages} {075106} (\bibinfo {year}
  {2014})}\BibitemShut {NoStop}%
\bibitem [{\citenamefont {Dijck}\ \emph {et~al.}(2023)\citenamefont {Dijck},
  \citenamefont {Warnecke}, \citenamefont {Wehrheim}, \citenamefont
  {Henninger}, \citenamefont {Eff}, \citenamefont {Georgiou}, \citenamefont
  {Graf}, \citenamefont {Kokh}, \citenamefont {Kozhiparambil~Sajith},
  \citenamefont {Mayo}, \citenamefont {Sch{\"a}fer}, \citenamefont {Volk},
  \citenamefont {Schmidt}, \citenamefont {Pfeifer},\ and\ \citenamefont
  {{Crespo L{\'o}pez-Urrutia}}}]{dijck_sympathetically_2023}%
  \BibitemOpen
  \bibfield  {author} {\bibinfo {author} {\bibfnamefont {E.~A.}\ \bibnamefont
  {Dijck}}, \bibinfo {author} {\bibfnamefont {C.}~\bibnamefont {Warnecke}},
  \bibinfo {author} {\bibfnamefont {M.}~\bibnamefont {Wehrheim}}, \bibinfo
  {author} {\bibfnamefont {R.~B.}\ \bibnamefont {Henninger}}, \bibinfo {author}
  {\bibfnamefont {J.}~\bibnamefont {Eff}}, \bibinfo {author} {\bibfnamefont
  {K.}~\bibnamefont {Georgiou}}, \bibinfo {author} {\bibfnamefont
  {A.}~\bibnamefont {Graf}}, \bibinfo {author} {\bibfnamefont {S.}~\bibnamefont
  {Kokh}}, \bibinfo {author} {\bibfnamefont {L.~P.}\ \bibnamefont
  {Kozhiparambil~Sajith}}, \bibinfo {author} {\bibfnamefont {C.}~\bibnamefont
  {Mayo}}, \bibinfo {author} {\bibfnamefont {V.~M.}\ \bibnamefont
  {Sch{\"a}fer}}, \bibinfo {author} {\bibfnamefont {C.}~\bibnamefont {Volk}},
  \bibinfo {author} {\bibfnamefont {P.~O.}\ \bibnamefont {Schmidt}}, \bibinfo
  {author} {\bibfnamefont {T.}~\bibnamefont {Pfeifer}},\ and\ \bibinfo {author}
  {\bibfnamefont {J.~R.}\ \bibnamefont {{Crespo L{\'o}pez-Urrutia}}},\ }\href
  {https://doi.org/10.1063/5.0160537} {\bibfield  {journal} {\bibinfo
  {journal} {Rev. Sci. Instrum.}\ }\textbf {\bibinfo {volume} {94}},\ \bibinfo
  {pages} {083203} (\bibinfo {year} {2023})}\BibitemShut {NoStop}%
\bibitem [{\citenamefont {Ji}\ \emph {et~al.}(2021)\citenamefont {Ji},
  \citenamefont {Zhou}, \citenamefont {Yan}, \citenamefont {He}, \citenamefont
  {Zhou}, \citenamefont {Barthwal}, \citenamefont {Yang}, \citenamefont {Duan},
  \citenamefont {Zhang}, \citenamefont {Xu}, \citenamefont {Wang},
  \citenamefont {Li}, \citenamefont {Gao}, \citenamefont {Chen}, \citenamefont
  {Wang},\ and\ \citenamefont {Zhan}}]{ji_actively_2021}%
  \BibitemOpen
  \bibfield  {author} {\bibinfo {author} {\bibfnamefont {Y.-H.}\ \bibnamefont
  {Ji}}, \bibinfo {author} {\bibfnamefont {L.}~\bibnamefont {Zhou}}, \bibinfo
  {author} {\bibfnamefont {S.-T.}\ \bibnamefont {Yan}}, \bibinfo {author}
  {\bibfnamefont {C.}~\bibnamefont {He}}, \bibinfo {author} {\bibfnamefont
  {C.}~\bibnamefont {Zhou}}, \bibinfo {author} {\bibfnamefont {S.}~\bibnamefont
  {Barthwal}}, \bibinfo {author} {\bibfnamefont {F.}~\bibnamefont {Yang}},
  \bibinfo {author} {\bibfnamefont {W.-T.}\ \bibnamefont {Duan}}, \bibinfo
  {author} {\bibfnamefont {W.-D.}\ \bibnamefont {Zhang}}, \bibinfo {author}
  {\bibfnamefont {R.-D.}\ \bibnamefont {Xu}}, \bibinfo {author} {\bibfnamefont
  {Q.}~\bibnamefont {Wang}}, \bibinfo {author} {\bibfnamefont {D.-X.}\
  \bibnamefont {Li}}, \bibinfo {author} {\bibfnamefont {J.-H.}\ \bibnamefont
  {Gao}}, \bibinfo {author} {\bibfnamefont {X.}~\bibnamefont {Chen}}, \bibinfo
  {author} {\bibfnamefont {J.}~\bibnamefont {Wang}},\ and\ \bibinfo {author}
  {\bibfnamefont {M.-S.}\ \bibnamefont {Zhan}},\ }\href
  {https://doi.org/10.1063/5.0053971} {\bibfield  {journal} {\bibinfo
  {journal} {Rev. Sci. Instrum.}\ }\textbf {\bibinfo {volume} {92}},\ \bibinfo
  {pages} {083201} (\bibinfo {year} {2021})}\BibitemShut {NoStop}%
\bibitem [{\citenamefont {Leopold}\ \emph {et~al.}(2019)\citenamefont
  {Leopold}, \citenamefont {King}, \citenamefont {Micke}, \citenamefont
  {{Bautista-Salvador}}, \citenamefont {Heip}, \citenamefont {Ospelkaus},
  \citenamefont {{Crespo L{\'o}pez-Urrutia}},\ and\ \citenamefont
  {Schmidt}}]{leopold_cryogenic_2019}%
  \BibitemOpen
  \bibfield  {author} {\bibinfo {author} {\bibfnamefont {T.}~\bibnamefont
  {Leopold}}, \bibinfo {author} {\bibfnamefont {S.~A.}\ \bibnamefont {King}},
  \bibinfo {author} {\bibfnamefont {P.}~\bibnamefont {Micke}}, \bibinfo
  {author} {\bibfnamefont {A.}~\bibnamefont {{Bautista-Salvador}}}, \bibinfo
  {author} {\bibfnamefont {J.~C.}\ \bibnamefont {Heip}}, \bibinfo {author}
  {\bibfnamefont {C.}~\bibnamefont {Ospelkaus}}, \bibinfo {author}
  {\bibfnamefont {J.~R.}\ \bibnamefont {{Crespo L{\'o}pez-Urrutia}}},\ and\
  \bibinfo {author} {\bibfnamefont {P.~O.}\ \bibnamefont {Schmidt}},\ }\href
  {https://doi.org/10.1063/1.5100594} {\bibfield  {journal} {\bibinfo
  {journal} {Rev. Sci. Instrum.}\ }\textbf {\bibinfo {volume} {90}},\ \bibinfo
  {pages} {073201} (\bibinfo {year} {2019})}\BibitemShut {NoStop}%
\bibitem [{\citenamefont {Farolfi}\ \emph {et~al.}(2019)\citenamefont
  {Farolfi}, \citenamefont {Trypogeorgos}, \citenamefont {Colzi}, \citenamefont
  {Fava}, \citenamefont {Lamporesi},\ and\ \citenamefont
  {Ferrari}}]{farolfi_design_2019}%
  \BibitemOpen
  \bibfield  {author} {\bibinfo {author} {\bibfnamefont {A.}~\bibnamefont
  {Farolfi}}, \bibinfo {author} {\bibfnamefont {D.}~\bibnamefont
  {Trypogeorgos}}, \bibinfo {author} {\bibfnamefont {G.}~\bibnamefont {Colzi}},
  \bibinfo {author} {\bibfnamefont {E.}~\bibnamefont {Fava}}, \bibinfo {author}
  {\bibfnamefont {G.}~\bibnamefont {Lamporesi}},\ and\ \bibinfo {author}
  {\bibfnamefont {G.}~\bibnamefont {Ferrari}},\ }\href
  {https://doi.org/10.1063/1.5119915} {\bibfield  {journal} {\bibinfo
  {journal} {Rev. Sci. Instrum.}\ }\textbf {\bibinfo {volume} {90}},\ \bibinfo
  {pages} {115114} (\bibinfo {year} {2019})}\BibitemShut {NoStop}%
\bibitem [{\citenamefont {Devlin}\ \emph {et~al.}(2019)\citenamefont {Devlin},
  \citenamefont {Wursten}, \citenamefont {Harrington}, \citenamefont {Higuchi},
  \citenamefont {Blessing}, \citenamefont {Borchert}, \citenamefont {Erlewein},
  \citenamefont {Hansen}, \citenamefont {Morgner}, \citenamefont {Bohman},
  \citenamefont {Mooser}, \citenamefont {Smorra}, \citenamefont {Wiesinger},
  \citenamefont {Blaum}, \citenamefont {Matsuda}, \citenamefont {Ospelkaus},
  \citenamefont {Quint}, \citenamefont {Walz}, \citenamefont {Yamazaki},\ and\
  \citenamefont {Ulmer}}]{devlin_superconducting_2019}%
  \BibitemOpen
  \bibfield  {author} {\bibinfo {author} {\bibfnamefont {J.~A.}\ \bibnamefont
  {Devlin}}, \bibinfo {author} {\bibfnamefont {E.}~\bibnamefont {Wursten}},
  \bibinfo {author} {\bibfnamefont {J.~A.}\ \bibnamefont {Harrington}},
  \bibinfo {author} {\bibfnamefont {T.}~\bibnamefont {Higuchi}}, \bibinfo
  {author} {\bibfnamefont {P.~E.}\ \bibnamefont {Blessing}}, \bibinfo {author}
  {\bibfnamefont {M.~J.}\ \bibnamefont {Borchert}}, \bibinfo {author}
  {\bibfnamefont {S.}~\bibnamefont {Erlewein}}, \bibinfo {author}
  {\bibfnamefont {J.~J.}\ \bibnamefont {Hansen}}, \bibinfo {author}
  {\bibfnamefont {J.}~\bibnamefont {Morgner}}, \bibinfo {author} {\bibfnamefont
  {M.~A.}\ \bibnamefont {Bohman}}, \bibinfo {author} {\bibfnamefont {A.~H.}\
  \bibnamefont {Mooser}}, \bibinfo {author} {\bibfnamefont {C.}~\bibnamefont
  {Smorra}}, \bibinfo {author} {\bibfnamefont {M.}~\bibnamefont {Wiesinger}},
  \bibinfo {author} {\bibfnamefont {K.}~\bibnamefont {Blaum}}, \bibinfo
  {author} {\bibfnamefont {Y.}~\bibnamefont {Matsuda}}, \bibinfo {author}
  {\bibfnamefont {C.}~\bibnamefont {Ospelkaus}}, \bibinfo {author}
  {\bibfnamefont {W.}~\bibnamefont {Quint}}, \bibinfo {author} {\bibfnamefont
  {J.}~\bibnamefont {Walz}}, \bibinfo {author} {\bibfnamefont {Y.}~\bibnamefont
  {Yamazaki}},\ and\ \bibinfo {author} {\bibfnamefont {S.}~\bibnamefont
  {Ulmer}},\ }\href {https://doi.org/10.1103/PhysRevApplied.12.044012}
  {\bibfield  {journal} {\bibinfo  {journal} {Phys. Rev. Applied}\ }\textbf
  {\bibinfo {volume} {12}},\ \bibinfo {pages} {044012} (\bibinfo {year}
  {2019})}\BibitemShut {NoStop}%
\bibitem [{\citenamefont {Wodey}\ \emph {et~al.}(2020)\citenamefont {Wodey},
  \citenamefont {Tell}, \citenamefont {Rasel}, \citenamefont {Schlippert},
  \citenamefont {Baur}, \citenamefont {Kissling}, \citenamefont {K{\"o}lliker},
  \citenamefont {Lorenz}, \citenamefont {Marrer}, \citenamefont
  {Schl{\"a}pfer}, \citenamefont {Widmer}, \citenamefont {Ufrecht},
  \citenamefont {Stuiber},\ and\ \citenamefont
  {Fierlinger}}]{wodey_scalable_2020}%
  \BibitemOpen
  \bibfield  {author} {\bibinfo {author} {\bibfnamefont {E.}~\bibnamefont
  {Wodey}}, \bibinfo {author} {\bibfnamefont {D.}~\bibnamefont {Tell}},
  \bibinfo {author} {\bibfnamefont {E.~M.}\ \bibnamefont {Rasel}}, \bibinfo
  {author} {\bibfnamefont {D.}~\bibnamefont {Schlippert}}, \bibinfo {author}
  {\bibfnamefont {R.}~\bibnamefont {Baur}}, \bibinfo {author} {\bibfnamefont
  {U.}~\bibnamefont {Kissling}}, \bibinfo {author} {\bibfnamefont
  {B.}~\bibnamefont {K{\"o}lliker}}, \bibinfo {author} {\bibfnamefont
  {M.}~\bibnamefont {Lorenz}}, \bibinfo {author} {\bibfnamefont
  {M.}~\bibnamefont {Marrer}}, \bibinfo {author} {\bibfnamefont
  {U.}~\bibnamefont {Schl{\"a}pfer}}, \bibinfo {author} {\bibfnamefont
  {M.}~\bibnamefont {Widmer}}, \bibinfo {author} {\bibfnamefont
  {C.}~\bibnamefont {Ufrecht}}, \bibinfo {author} {\bibfnamefont
  {S.}~\bibnamefont {Stuiber}},\ and\ \bibinfo {author} {\bibfnamefont
  {P.}~\bibnamefont {Fierlinger}},\ }\href {https://doi.org/10.1063/1.5141340}
  {\bibfield  {journal} {\bibinfo  {journal} {Rev. Sci. Instrum.}\ }\textbf
  {\bibinfo {volume} {91}},\ \bibinfo {pages} {035117} (\bibinfo {year}
  {2020})}\BibitemShut {NoStop}%
\bibitem [{\citenamefont {Vovrosh}\ \emph {et~al.}(2018)\citenamefont
  {Vovrosh}, \citenamefont {Voulazeris}, \citenamefont {Petrov}, \citenamefont
  {Zou}, \citenamefont {Gaber}, \citenamefont {Benn}, \citenamefont {Woolger},
  \citenamefont {Attallah}, \citenamefont {Boyer}, \citenamefont {Bongs},\ and\
  \citenamefont {Holynski}}]{vovrosh_additive_2018}%
  \BibitemOpen
  \bibfield  {author} {\bibinfo {author} {\bibfnamefont {J.}~\bibnamefont
  {Vovrosh}}, \bibinfo {author} {\bibfnamefont {G.}~\bibnamefont {Voulazeris}},
  \bibinfo {author} {\bibfnamefont {P.~G.}\ \bibnamefont {Petrov}}, \bibinfo
  {author} {\bibfnamefont {J.}~\bibnamefont {Zou}}, \bibinfo {author}
  {\bibfnamefont {Y.}~\bibnamefont {Gaber}}, \bibinfo {author} {\bibfnamefont
  {L.}~\bibnamefont {Benn}}, \bibinfo {author} {\bibfnamefont {D.}~\bibnamefont
  {Woolger}}, \bibinfo {author} {\bibfnamefont {M.~M.}\ \bibnamefont
  {Attallah}}, \bibinfo {author} {\bibfnamefont {V.}~\bibnamefont {Boyer}},
  \bibinfo {author} {\bibfnamefont {K.}~\bibnamefont {Bongs}},\ and\ \bibinfo
  {author} {\bibfnamefont {M.}~\bibnamefont {Holynski}},\ }\href
  {https://doi.org/10.1038/s41598-018-20352-x} {\bibfield  {journal} {\bibinfo
  {journal} {Sci. Rep.}\ }\textbf {\bibinfo {volume} {8}},\ \bibinfo {pages}
  {2023} (\bibinfo {year} {2018})}\BibitemShut {NoStop}%
\bibitem [{\citenamefont {Brandl}\ \emph {et~al.}(2016)\citenamefont {Brandl},
  \citenamefont {van Mourik}, \citenamefont {Postler}, \citenamefont {Nolf},
  \citenamefont {Lakhmanskiy}, \citenamefont {Paiva}, \citenamefont
  {M{\"o}ller}, \citenamefont {Daniilidis}, \citenamefont {H{\"a}ffner},
  \citenamefont {Kaushal}, \citenamefont {Ruster}, \citenamefont
  {Warschburger}, \citenamefont {Kaufmann}, \citenamefont {Poschinger},
  \citenamefont {{Schmidt-Kaler}}, \citenamefont {Schindler}, \citenamefont
  {Monz},\ and\ \citenamefont {Blatt}}]{brandl_cryogenic_2016-1}%
  \BibitemOpen
  \bibfield  {author} {\bibinfo {author} {\bibfnamefont {M.~F.}\ \bibnamefont
  {Brandl}}, \bibinfo {author} {\bibfnamefont {M.~W.}\ \bibnamefont {van
  Mourik}}, \bibinfo {author} {\bibfnamefont {L.}~\bibnamefont {Postler}},
  \bibinfo {author} {\bibfnamefont {A.}~\bibnamefont {Nolf}}, \bibinfo {author}
  {\bibfnamefont {K.}~\bibnamefont {Lakhmanskiy}}, \bibinfo {author}
  {\bibfnamefont {R.~R.}\ \bibnamefont {Paiva}}, \bibinfo {author}
  {\bibfnamefont {S.}~\bibnamefont {M{\"o}ller}}, \bibinfo {author}
  {\bibfnamefont {N.}~\bibnamefont {Daniilidis}}, \bibinfo {author}
  {\bibfnamefont {H.}~\bibnamefont {H{\"a}ffner}}, \bibinfo {author}
  {\bibfnamefont {V.}~\bibnamefont {Kaushal}}, \bibinfo {author} {\bibfnamefont
  {T.}~\bibnamefont {Ruster}}, \bibinfo {author} {\bibfnamefont
  {C.}~\bibnamefont {Warschburger}}, \bibinfo {author} {\bibfnamefont
  {H.}~\bibnamefont {Kaufmann}}, \bibinfo {author} {\bibfnamefont {U.~G.}\
  \bibnamefont {Poschinger}}, \bibinfo {author} {\bibfnamefont
  {F.}~\bibnamefont {{Schmidt-Kaler}}}, \bibinfo {author} {\bibfnamefont
  {P.}~\bibnamefont {Schindler}}, \bibinfo {author} {\bibfnamefont
  {T.}~\bibnamefont {Monz}},\ and\ \bibinfo {author} {\bibfnamefont
  {R.}~\bibnamefont {Blatt}},\ }\href {https://doi.org/10.1063/1.4966970}
  {\bibfield  {journal} {\bibinfo  {journal} {Rev. Sci. Instrum.}\ }\textbf
  {\bibinfo {volume} {87}},\ \bibinfo {pages} {113103} (\bibinfo {year}
  {2016})}\BibitemShut {NoStop}%
\bibitem [{\citenamefont {Altarev}\ \emph {et~al.}(2015)\citenamefont
  {Altarev}, \citenamefont {Bales}, \citenamefont {Beck}, \citenamefont
  {Chupp}, \citenamefont {Fierlinger}, \citenamefont {Fierlinger},
  \citenamefont {Kuchler}, \citenamefont {Lins}, \citenamefont {Marino},
  \citenamefont {Niessen}, \citenamefont {Petzoldt}, \citenamefont
  {Schl{\"a}pfer}, \citenamefont {Schnabel}, \citenamefont {Singh},
  \citenamefont {Stoepler}, \citenamefont {Stuiber}, \citenamefont {Sturm},
  \citenamefont {Taubenheim},\ and\ \citenamefont
  {Voigt}}]{altarev_large-scale_2015}%
  \BibitemOpen
  \bibfield  {author} {\bibinfo {author} {\bibfnamefont {I.}~\bibnamefont
  {Altarev}}, \bibinfo {author} {\bibfnamefont {M.}~\bibnamefont {Bales}},
  \bibinfo {author} {\bibfnamefont {D.~H.}\ \bibnamefont {Beck}}, \bibinfo
  {author} {\bibfnamefont {T.}~\bibnamefont {Chupp}}, \bibinfo {author}
  {\bibfnamefont {K.}~\bibnamefont {Fierlinger}}, \bibinfo {author}
  {\bibfnamefont {P.}~\bibnamefont {Fierlinger}}, \bibinfo {author}
  {\bibfnamefont {F.}~\bibnamefont {Kuchler}}, \bibinfo {author} {\bibfnamefont
  {T.}~\bibnamefont {Lins}}, \bibinfo {author} {\bibfnamefont {M.~G.}\
  \bibnamefont {Marino}}, \bibinfo {author} {\bibfnamefont {B.}~\bibnamefont
  {Niessen}}, \bibinfo {author} {\bibfnamefont {G.}~\bibnamefont {Petzoldt}},
  \bibinfo {author} {\bibfnamefont {U.}~\bibnamefont {Schl{\"a}pfer}}, \bibinfo
  {author} {\bibfnamefont {A.}~\bibnamefont {Schnabel}}, \bibinfo {author}
  {\bibfnamefont {J.~T.}\ \bibnamefont {Singh}}, \bibinfo {author}
  {\bibfnamefont {R.}~\bibnamefont {Stoepler}}, \bibinfo {author}
  {\bibfnamefont {S.}~\bibnamefont {Stuiber}}, \bibinfo {author} {\bibfnamefont
  {M.}~\bibnamefont {Sturm}}, \bibinfo {author} {\bibfnamefont
  {B.}~\bibnamefont {Taubenheim}},\ and\ \bibinfo {author} {\bibfnamefont
  {J.}~\bibnamefont {Voigt}},\ }\href {https://doi.org/10.1063/1.4919366}
  {\bibfield  {journal} {\bibinfo  {journal} {J. Appl. Phys.}\ }\textbf
  {\bibinfo {volume} {117}},\ \bibinfo {pages} {183903} (\bibinfo {year}
  {2015})}\BibitemShut {NoStop}%
\bibitem [{\citenamefont {Lee}\ and\ \citenamefont
  {Romalis}(2008)}]{lee_calculation_2008}%
  \BibitemOpen
  \bibfield  {author} {\bibinfo {author} {\bibfnamefont {S.-K.}\ \bibnamefont
  {Lee}}\ and\ \bibinfo {author} {\bibfnamefont {M.~V.}\ \bibnamefont
  {Romalis}},\ }\href {https://doi.org/10.1063/1.2885711} {\bibfield  {journal}
  {\bibinfo  {journal} {J. Appl. Phys.}\ }\textbf {\bibinfo {volume} {103}},\
  \bibinfo {pages} {084904} (\bibinfo {year} {2008})}\BibitemShut {NoStop}%
\bibitem [{\citenamefont {Mager}(1970)}]{mager_magnetic_1970}%
  \BibitemOpen
  \bibfield  {author} {\bibinfo {author} {\bibfnamefont {A.}~\bibnamefont
  {Mager}},\ }\href {https://doi.org/10.1109/TMAG.1970.1066714} {\bibfield
  {journal} {\bibinfo  {journal} {IEEE Trans. Magn.}\ }\textbf {\bibinfo
  {volume} {6}},\ \bibinfo {pages} {67} (\bibinfo {year} {1970})}\BibitemShut
  {NoStop}%
\bibitem [{\citenamefont {Hanson}\ and\ \citenamefont
  {Pipkin}(1965)}]{hanson_magnetically_1965}%
  \BibitemOpen
  \bibfield  {author} {\bibinfo {author} {\bibfnamefont {R.~J.}\ \bibnamefont
  {Hanson}}\ and\ \bibinfo {author} {\bibfnamefont {F.~M.}\ \bibnamefont
  {Pipkin}},\ }\href {https://doi.org/10.1063/1.1719514} {\bibfield  {journal}
  {\bibinfo  {journal} {Rev. Sci. Instrum.}\ }\textbf {\bibinfo {volume}
  {36}},\ \bibinfo {pages} {179} (\bibinfo {year} {1965})}\BibitemShut
  {NoStop}%
\bibitem [{\citenamefont {Ruster}\ \emph {et~al.}(2016)\citenamefont {Ruster},
  \citenamefont {Schmiegelow}, \citenamefont {Kaufmann}, \citenamefont
  {Warschburger}, \citenamefont {{Schmidt-Kaler}},\ and\ \citenamefont
  {Poschinger}}]{ruster_long-lived_2016}%
  \BibitemOpen
  \bibfield  {author} {\bibinfo {author} {\bibfnamefont {T.}~\bibnamefont
  {Ruster}}, \bibinfo {author} {\bibfnamefont {C.~T.}\ \bibnamefont
  {Schmiegelow}}, \bibinfo {author} {\bibfnamefont {H.}~\bibnamefont
  {Kaufmann}}, \bibinfo {author} {\bibfnamefont {C.}~\bibnamefont
  {Warschburger}}, \bibinfo {author} {\bibfnamefont {F.}~\bibnamefont
  {{Schmidt-Kaler}}},\ and\ \bibinfo {author} {\bibfnamefont {U.~G.}\
  \bibnamefont {Poschinger}},\ }\href
  {https://doi.org/10.1007/s00340-016-6527-4} {\bibfield  {journal} {\bibinfo
  {journal} {Appl. Phys. B}\ }\textbf {\bibinfo {volume} {122}},\ \bibinfo
  {pages} {254} (\bibinfo {year} {2016})}\BibitemShut {NoStop}%
\bibitem [{\citenamefont {Dub{\'e}}\ \emph {et~al.}(2005)\citenamefont
  {Dub{\'e}}, \citenamefont {Madej}, \citenamefont {Bernard}, \citenamefont
  {Marmet}, \citenamefont {Boulanger},\ and\ \citenamefont
  {Cundy}}]{dube_electric_2005}%
  \BibitemOpen
  \bibfield  {author} {\bibinfo {author} {\bibfnamefont {P.}~\bibnamefont
  {Dub{\'e}}}, \bibinfo {author} {\bibfnamefont {A.}~\bibnamefont {Madej}},
  \bibinfo {author} {\bibfnamefont {J.}~\bibnamefont {Bernard}}, \bibinfo
  {author} {\bibfnamefont {L.}~\bibnamefont {Marmet}}, \bibinfo {author}
  {\bibfnamefont {J.-S.}\ \bibnamefont {Boulanger}},\ and\ \bibinfo {author}
  {\bibfnamefont {S.}~\bibnamefont {Cundy}},\ }\href
  {https://doi.org/10.1103/PhysRevLett.95.033001} {\bibfield  {journal}
  {\bibinfo  {journal} {Phys. Rev. Lett.}\ }\textbf {\bibinfo {volume} {95}},\
  \bibinfo {pages} {033001} (\bibinfo {year} {2005})}\BibitemShut {NoStop}%
\bibitem [{\citenamefont {Margolis}\ \emph {et~al.}(2004)\citenamefont
  {Margolis}, \citenamefont {Barwood}, \citenamefont {Huang}, \citenamefont
  {Klein}, \citenamefont {Lea}, \citenamefont {Szymaniec},\ and\ \citenamefont
  {Gill}}]{margolis_hertz-level_2004}%
  \BibitemOpen
  \bibfield  {author} {\bibinfo {author} {\bibfnamefont {H.~S.}\ \bibnamefont
  {Margolis}}, \bibinfo {author} {\bibfnamefont {G.~P.}\ \bibnamefont
  {Barwood}}, \bibinfo {author} {\bibfnamefont {G.}~\bibnamefont {Huang}},
  \bibinfo {author} {\bibfnamefont {H.~A.}\ \bibnamefont {Klein}}, \bibinfo
  {author} {\bibfnamefont {S.~N.}\ \bibnamefont {Lea}}, \bibinfo {author}
  {\bibfnamefont {K.}~\bibnamefont {Szymaniec}},\ and\ \bibinfo {author}
  {\bibfnamefont {P.}~\bibnamefont {Gill}},\ }\href
  {https://doi.org/10.1126/science.1105497} {\bibfield  {journal} {\bibinfo
  {journal} {Science}\ }\textbf {\bibinfo {volume} {306}},\ \bibinfo {pages}
  {1355} (\bibinfo {year} {2004})}\BibitemShut {NoStop}%
\bibitem [{\citenamefont {Chwalla}\ \emph {et~al.}(2009)\citenamefont
  {Chwalla}, \citenamefont {Benhelm}, \citenamefont {Kim}, \citenamefont
  {Kirchmair}, \citenamefont {Monz}, \citenamefont {Riebe}, \citenamefont
  {Schindler}, \citenamefont {Villar}, \citenamefont {H{\"a}nsel},
  \citenamefont {Roos}, \citenamefont {Blatt}, \citenamefont {Abgrall},
  \citenamefont {Santarelli}, \citenamefont {Rovera},\ and\ \citenamefont
  {Laurent}}]{chwalla_absolute_2009}%
  \BibitemOpen
  \bibfield  {author} {\bibinfo {author} {\bibfnamefont {M.}~\bibnamefont
  {Chwalla}}, \bibinfo {author} {\bibfnamefont {J.}~\bibnamefont {Benhelm}},
  \bibinfo {author} {\bibfnamefont {K.}~\bibnamefont {Kim}}, \bibinfo {author}
  {\bibfnamefont {G.}~\bibnamefont {Kirchmair}}, \bibinfo {author}
  {\bibfnamefont {T.}~\bibnamefont {Monz}}, \bibinfo {author} {\bibfnamefont
  {M.}~\bibnamefont {Riebe}}, \bibinfo {author} {\bibfnamefont
  {P.}~\bibnamefont {Schindler}}, \bibinfo {author} {\bibfnamefont
  {A.}~\bibnamefont {Villar}}, \bibinfo {author} {\bibfnamefont
  {W.}~\bibnamefont {H{\"a}nsel}}, \bibinfo {author} {\bibfnamefont
  {C.}~\bibnamefont {Roos}}, \bibinfo {author} {\bibfnamefont {R.}~\bibnamefont
  {Blatt}}, \bibinfo {author} {\bibfnamefont {M.}~\bibnamefont {Abgrall}},
  \bibinfo {author} {\bibfnamefont {G.}~\bibnamefont {Santarelli}}, \bibinfo
  {author} {\bibfnamefont {G.}~\bibnamefont {Rovera}},\ and\ \bibinfo {author}
  {\bibfnamefont {{\relax Ph}.}~\bibnamefont {Laurent}},\ }\href
  {https://doi.org/10.1103/PhysRevLett.102.023002} {\bibfield  {journal}
  {\bibinfo  {journal} {Phys. Rev. Lett.}\ }\textbf {\bibinfo {volume} {102}},\
  \bibinfo {pages} {023002} (\bibinfo {year} {2009})}\BibitemShut {NoStop}%
\bibitem [{\citenamefont {Dub{\'e}}\ \emph {et~al.}(2013)\citenamefont
  {Dub{\'e}}, \citenamefont {Madej}, \citenamefont {Zhou},\ and\ \citenamefont
  {Bernard}}]{dube_evaluation_2013}%
  \BibitemOpen
  \bibfield  {author} {\bibinfo {author} {\bibfnamefont {P.}~\bibnamefont
  {Dub{\'e}}}, \bibinfo {author} {\bibfnamefont {A.~A.}\ \bibnamefont {Madej}},
  \bibinfo {author} {\bibfnamefont {Z.}~\bibnamefont {Zhou}},\ and\ \bibinfo
  {author} {\bibfnamefont {J.~E.}\ \bibnamefont {Bernard}},\ }\href
  {https://doi.org/10.1103/PhysRevA.87.023806} {\bibfield  {journal} {\bibinfo
  {journal} {Phys. Rev. A}\ }\textbf {\bibinfo {volume} {87}},\ \bibinfo
  {pages} {023806} (\bibinfo {year} {2013})}\BibitemShut {NoStop}%
\bibitem [{\citenamefont {Oskay}\ \emph {et~al.}(2005)\citenamefont {Oskay},
  \citenamefont {Itano},\ and\ \citenamefont
  {Bergquist}}]{oskay_measurement_2005}%
  \BibitemOpen
  \bibfield  {author} {\bibinfo {author} {\bibfnamefont {W.}~\bibnamefont
  {Oskay}}, \bibinfo {author} {\bibfnamefont {W.}~\bibnamefont {Itano}},\ and\
  \bibinfo {author} {\bibfnamefont {J.}~\bibnamefont {Bergquist}},\ }\href
  {https://doi.org/10.1103/PhysRevLett.94.163001} {\bibfield  {journal}
  {\bibinfo  {journal} {Phys. Rev. Lett.}\ }\textbf {\bibinfo {volume} {94}},\
  \bibinfo {pages} {163001} (\bibinfo {year} {2005})}\BibitemShut {NoStop}%
\bibitem [{\citenamefont {Schneider}\ \emph {et~al.}(2005)\citenamefont
  {Schneider}, \citenamefont {Peik},\ and\ \citenamefont
  {Tamm}}]{schneider_sub-hertz_2005}%
  \BibitemOpen
  \bibfield  {author} {\bibinfo {author} {\bibfnamefont {T.}~\bibnamefont
  {Schneider}}, \bibinfo {author} {\bibfnamefont {E.}~\bibnamefont {Peik}},\
  and\ \bibinfo {author} {\bibfnamefont {{\relax Chr}.}~\bibnamefont {Tamm}},\
  }\href {https://doi.org/10.1103/PhysRevLett.94.230801} {\bibfield  {journal}
  {\bibinfo  {journal} {Phys. Rev. Lett.}\ }\textbf {\bibinfo {volume} {94}},\
  \bibinfo {pages} {230801} (\bibinfo {year} {2005})}\BibitemShut {NoStop}%
\bibitem [{\citenamefont {Tan}\ \emph {et~al.}(2019)\citenamefont {Tan},
  \citenamefont {Kaewuam}, \citenamefont {Arnold}, \citenamefont {Chanu},
  \citenamefont {Zhang}, \citenamefont {Safronova},\ and\ \citenamefont
  {Barrett}}]{tan_suppressing_2019}%
  \BibitemOpen
  \bibfield  {author} {\bibinfo {author} {\bibfnamefont {T.~R.}\ \bibnamefont
  {Tan}}, \bibinfo {author} {\bibfnamefont {R.}~\bibnamefont {Kaewuam}},
  \bibinfo {author} {\bibfnamefont {K.~J.}\ \bibnamefont {Arnold}}, \bibinfo
  {author} {\bibfnamefont {S.~R.}\ \bibnamefont {Chanu}}, \bibinfo {author}
  {\bibfnamefont {Z.}~\bibnamefont {Zhang}}, \bibinfo {author} {\bibfnamefont
  {M.~S.}\ \bibnamefont {Safronova}},\ and\ \bibinfo {author} {\bibfnamefont
  {M.~D.}\ \bibnamefont {Barrett}},\ }\href
  {https://doi.org/10.1103/PhysRevLett.123.063201} {\bibfield  {journal}
  {\bibinfo  {journal} {Phys. Rev. Lett.}\ }\textbf {\bibinfo {volume} {123}},\
  \bibinfo {pages} {063201} (\bibinfo {year} {2019})}\BibitemShut {NoStop}%
\bibitem [{\citenamefont {Steinel}\ \emph {et~al.}(2023)\citenamefont
  {Steinel}, \citenamefont {Shao}, \citenamefont {Filzinger}, \citenamefont
  {Lipphardt}, \citenamefont {Brinkmann}, \citenamefont {Didier}, \citenamefont
  {Mehlst{\"a}ubler}, \citenamefont {Lindvall}, \citenamefont {Peik},\ and\
  \citenamefont {Huntemann}}]{steinel_evaluation_2023}%
  \BibitemOpen
  \bibfield  {author} {\bibinfo {author} {\bibfnamefont {M.}~\bibnamefont
  {Steinel}}, \bibinfo {author} {\bibfnamefont {H.}~\bibnamefont {Shao}},
  \bibinfo {author} {\bibfnamefont {M.}~\bibnamefont {Filzinger}}, \bibinfo
  {author} {\bibfnamefont {B.}~\bibnamefont {Lipphardt}}, \bibinfo {author}
  {\bibfnamefont {M.}~\bibnamefont {Brinkmann}}, \bibinfo {author}
  {\bibfnamefont {A.}~\bibnamefont {Didier}}, \bibinfo {author} {\bibfnamefont
  {T.~E.}\ \bibnamefont {Mehlst{\"a}ubler}}, \bibinfo {author} {\bibfnamefont
  {T.}~\bibnamefont {Lindvall}}, \bibinfo {author} {\bibfnamefont
  {E.}~\bibnamefont {Peik}},\ and\ \bibinfo {author} {\bibfnamefont
  {N.}~\bibnamefont {Huntemann}},\ }\href
  {https://doi.org/10.1103/PhysRevLett.131.083002} {\bibfield  {journal}
  {\bibinfo  {journal} {Phys. Rev. Lett.}\ }\textbf {\bibinfo {volume} {131}},\
  \bibinfo {pages} {083002} (\bibinfo {year} {2023})}\BibitemShut {NoStop}%
\bibitem [{\citenamefont {Aharon}\ \emph {et~al.}(2019)\citenamefont {Aharon},
  \citenamefont {Spethmann}, \citenamefont {Leroux}, \citenamefont {Schmidt},\
  and\ \citenamefont {Retzker}}]{aharon_robust_2019}%
  \BibitemOpen
  \bibfield  {author} {\bibinfo {author} {\bibfnamefont {N.}~\bibnamefont
  {Aharon}}, \bibinfo {author} {\bibfnamefont {N.}~\bibnamefont {Spethmann}},
  \bibinfo {author} {\bibfnamefont {I.~D.}\ \bibnamefont {Leroux}}, \bibinfo
  {author} {\bibfnamefont {P.~O.}\ \bibnamefont {Schmidt}},\ and\ \bibinfo
  {author} {\bibfnamefont {A.}~\bibnamefont {Retzker}},\ }\href
  {https://doi.org/10.1088/1367-2630/ab3871} {\bibfield  {journal} {\bibinfo
  {journal} {New J. Phys.}\ }\textbf {\bibinfo {volume} {21}},\ \bibinfo
  {pages} {083040} (\bibinfo {year} {2019})}\BibitemShut {NoStop}%
\bibitem [{\citenamefont {Keller}\ \emph {et~al.}(2019)\citenamefont {Keller},
  \citenamefont {Burgermeister}, \citenamefont {Kalincev}, \citenamefont
  {Didier}, \citenamefont {Kulosa}, \citenamefont {Nordmann}, \citenamefont
  {Kiethe},\ and\ \citenamefont {Mehlst{\"a}ubler}}]{keller_controlling_2019}%
  \BibitemOpen
  \bibfield  {author} {\bibinfo {author} {\bibfnamefont {J.}~\bibnamefont
  {Keller}}, \bibinfo {author} {\bibfnamefont {T.}~\bibnamefont
  {Burgermeister}}, \bibinfo {author} {\bibfnamefont {D.}~\bibnamefont
  {Kalincev}}, \bibinfo {author} {\bibfnamefont {A.}~\bibnamefont {Didier}},
  \bibinfo {author} {\bibfnamefont {A.~P.}\ \bibnamefont {Kulosa}}, \bibinfo
  {author} {\bibfnamefont {T.}~\bibnamefont {Nordmann}}, \bibinfo {author}
  {\bibfnamefont {J.}~\bibnamefont {Kiethe}},\ and\ \bibinfo {author}
  {\bibfnamefont {T.~E.}\ \bibnamefont {Mehlst{\"a}ubler}},\ }\href
  {https://doi.org/10.1103/PhysRevA.99.013405} {\bibfield  {journal} {\bibinfo
  {journal} {Phys. Rev. A}\ }\textbf {\bibinfo {volume} {99}},\ \bibinfo
  {pages} {013405} (\bibinfo {year} {2019})}\BibitemShut {NoStop}%
\bibitem [{\citenamefont {Arnold}\ \emph {et~al.}(2015)\citenamefont {Arnold},
  \citenamefont {Hajiyev}, \citenamefont {Paez}, \citenamefont {Lee},
  \citenamefont {Barrett},\ and\ \citenamefont
  {Bollinger}}]{arnold_prospects_2015}%
  \BibitemOpen
  \bibfield  {author} {\bibinfo {author} {\bibfnamefont {K.}~\bibnamefont
  {Arnold}}, \bibinfo {author} {\bibfnamefont {E.}~\bibnamefont {Hajiyev}},
  \bibinfo {author} {\bibfnamefont {E.}~\bibnamefont {Paez}}, \bibinfo {author}
  {\bibfnamefont {C.~H.}\ \bibnamefont {Lee}}, \bibinfo {author} {\bibfnamefont
  {M.~D.}\ \bibnamefont {Barrett}},\ and\ \bibinfo {author} {\bibfnamefont
  {J.}~\bibnamefont {Bollinger}},\ }\href
  {https://doi.org/10.1103/PhysRevA.92.032108} {\bibfield  {journal} {\bibinfo
  {journal} {Phys. Rev. A}\ }\textbf {\bibinfo {volume} {92}},\ \bibinfo
  {pages} {032108} (\bibinfo {year} {2015})}\BibitemShut {NoStop}%
\bibitem [{\citenamefont {Herschbach}\ \emph {et~al.}(2012)\citenamefont
  {Herschbach}, \citenamefont {Pyka}, \citenamefont {Keller},\ and\
  \citenamefont {Mehlst{\"a}ubler}}]{herschbach_linear_2012}%
  \BibitemOpen
  \bibfield  {author} {\bibinfo {author} {\bibfnamefont {N.}~\bibnamefont
  {Herschbach}}, \bibinfo {author} {\bibfnamefont {K.}~\bibnamefont {Pyka}},
  \bibinfo {author} {\bibfnamefont {J.}~\bibnamefont {Keller}},\ and\ \bibinfo
  {author} {\bibfnamefont {T.~E.}\ \bibnamefont {Mehlst{\"a}ubler}},\ }\href
  {https://doi.org/10.1007/s00340-011-4790-y} {\bibfield  {journal} {\bibinfo
  {journal} {Appl. Phys. B}\ }\textbf {\bibinfo {volume} {107}},\ \bibinfo
  {pages} {891} (\bibinfo {year} {2012})}\BibitemShut {NoStop}%
\bibitem [{\citenamefont {Champenois}\ \emph {et~al.}(2010)\citenamefont
  {Champenois}, \citenamefont {Marciante}, \citenamefont
  {{Pedregosa-Gutierrez}}, \citenamefont {Houssin},\ and\ \citenamefont
  {Knoop}}]{champenois_ion_2010}%
  \BibitemOpen
  \bibfield  {author} {\bibinfo {author} {\bibfnamefont {C.}~\bibnamefont
  {Champenois}}, \bibinfo {author} {\bibfnamefont {M.}~\bibnamefont
  {Marciante}}, \bibinfo {author} {\bibfnamefont {J.}~\bibnamefont
  {{Pedregosa-Gutierrez}}}, \bibinfo {author} {\bibfnamefont {M.}~\bibnamefont
  {Houssin}},\ and\ \bibinfo {author} {\bibfnamefont {M.}~\bibnamefont
  {Knoop}},\ }\bibfield  {journal} {\bibinfo  {journal} {Phys. Rev. A}\
  }\textbf {\bibinfo {volume} {81}},\ \href
  {https://doi.org/10.1103/PhysRevA.81.043410} {10.1103/PhysRevA.81.043410}
  (\bibinfo {year} {2010})\BibitemShut {NoStop}%
\bibitem [{\citenamefont {Hahn}(1950)}]{hahn_spin_1950}%
  \BibitemOpen
  \bibfield  {author} {\bibinfo {author} {\bibfnamefont {E.~L.}\ \bibnamefont
  {Hahn}},\ }\href {https://doi.org/10.1103/PhysRev.80.580} {\bibfield
  {journal} {\bibinfo  {journal} {Phys. Rev.}\ }\textbf {\bibinfo {volume}
  {80}},\ \bibinfo {pages} {580} (\bibinfo {year} {1950})}\BibitemShut
  {NoStop}%
\bibitem [{\citenamefont {Viola}\ \emph {et~al.}(1999)\citenamefont {Viola},
  \citenamefont {Knill},\ and\ \citenamefont {Lloyd}}]{viola_dynamical_1999}%
  \BibitemOpen
  \bibfield  {author} {\bibinfo {author} {\bibfnamefont {L.}~\bibnamefont
  {Viola}}, \bibinfo {author} {\bibfnamefont {E.}~\bibnamefont {Knill}},\ and\
  \bibinfo {author} {\bibfnamefont {S.}~\bibnamefont {Lloyd}},\ }\href
  {https://doi.org/10.1103/PhysRevLett.82.2417} {\bibfield  {journal} {\bibinfo
   {journal} {Phys. Rev. Lett.}\ }\textbf {\bibinfo {volume} {82}},\ \bibinfo
  {pages} {2417} (\bibinfo {year} {1999})}\BibitemShut {NoStop}%
\bibitem [{\citenamefont {Valahu}\ \emph {et~al.}(2022)\citenamefont {Valahu},
  \citenamefont {Apostolatos}, \citenamefont {Weidt},\ and\ \citenamefont
  {Hensinger}}]{valahu_quantum_2022}%
  \BibitemOpen
  \bibfield  {author} {\bibinfo {author} {\bibfnamefont {C.~H.}\ \bibnamefont
  {Valahu}}, \bibinfo {author} {\bibfnamefont {I.}~\bibnamefont {Apostolatos}},
  \bibinfo {author} {\bibfnamefont {S.}~\bibnamefont {Weidt}},\ and\ \bibinfo
  {author} {\bibfnamefont {W.~K.}\ \bibnamefont {Hensinger}},\ }\href
  {https://doi.org/10.1088/1361-6455/ac8eff} {\bibfield  {journal} {\bibinfo
  {journal} {J. Phys. B: At. Mol. Opt. Phys.}\ }\textbf {\bibinfo {volume}
  {55}},\ \bibinfo {pages} {204003} (\bibinfo {year} {2022})}\BibitemShut
  {NoStop}%
\bibitem [{\citenamefont {Cai}\ and\ \citenamefont
  {Xia}(2022)}]{cai_optimizing_2022}%
  \BibitemOpen
  \bibfield  {author} {\bibinfo {author} {\bibfnamefont {M.}~\bibnamefont
  {Cai}}\ and\ \bibinfo {author} {\bibfnamefont {K.}~\bibnamefont {Xia}},\
  }\href {https://doi.org/10.1103/PhysRevA.106.042434} {\bibfield  {journal}
  {\bibinfo  {journal} {Phys. Rev. A}\ }\textbf {\bibinfo {volume} {106}},\
  \bibinfo {pages} {042434} (\bibinfo {year} {2022})}\BibitemShut {NoStop}%
\bibitem [{\citenamefont {Yal{\c c}{\i}nkaya}\ \emph
  {et~al.}(2019)\citenamefont {Yal{\c c}{\i}nkaya}, \citenamefont {{\c
  C}akmak}, \citenamefont {Karpat},\ and\ \citenamefont
  {Fanchini}}]{yalcinkaya_continuous_2019}%
  \BibitemOpen
  \bibfield  {author} {\bibinfo {author} {\bibfnamefont {{\.I}.}~\bibnamefont
  {Yal{\c c}{\i}nkaya}}, \bibinfo {author} {\bibfnamefont {B.}~\bibnamefont
  {{\c C}akmak}}, \bibinfo {author} {\bibfnamefont {G.}~\bibnamefont
  {Karpat}},\ and\ \bibinfo {author} {\bibfnamefont {F.~F.}\ \bibnamefont
  {Fanchini}},\ }\href {https://doi.org/10.1007/s11128-019-2271-0} {\bibfield
  {journal} {\bibinfo  {journal} {Quantum Inf. Process.}\ }\textbf {\bibinfo
  {volume} {18}},\ \bibinfo {pages} {156} (\bibinfo {year} {2019})}\BibitemShut
  {NoStop}%
\bibitem [{\citenamefont {Cohen}\ \emph {et~al.}(2017)\citenamefont {Cohen},
  \citenamefont {Aharon},\ and\ \citenamefont
  {Retzker}}]{cohen_continuous_2017}%
  \BibitemOpen
  \bibfield  {author} {\bibinfo {author} {\bibfnamefont {I.}~\bibnamefont
  {Cohen}}, \bibinfo {author} {\bibfnamefont {N.}~\bibnamefont {Aharon}},\ and\
  \bibinfo {author} {\bibfnamefont {A.}~\bibnamefont {Retzker}},\ }\href
  {https://doi.org/10.1002/prop.201600071} {\bibfield  {journal} {\bibinfo
  {journal} {Fortschritte Phys.}\ }\textbf {\bibinfo {volume} {65}},\ \bibinfo
  {pages} {1600071} (\bibinfo {year} {2017})}\BibitemShut {NoStop}%
\bibitem [{\citenamefont {Mikelsons}\ \emph {et~al.}(2015)\citenamefont
  {Mikelsons}, \citenamefont {Cohen}, \citenamefont {Retzker},\ and\
  \citenamefont {Plenio}}]{mikelsons_universal_2015}%
  \BibitemOpen
  \bibfield  {author} {\bibinfo {author} {\bibfnamefont {G.}~\bibnamefont
  {Mikelsons}}, \bibinfo {author} {\bibfnamefont {I.}~\bibnamefont {Cohen}},
  \bibinfo {author} {\bibfnamefont {A.}~\bibnamefont {Retzker}},\ and\ \bibinfo
  {author} {\bibfnamefont {M.~B.}\ \bibnamefont {Plenio}},\ }\href
  {https://doi.org/10.1088/1367-2630/17/5/053032} {\bibfield  {journal}
  {\bibinfo  {journal} {New J. Phys.}\ }\textbf {\bibinfo {volume} {17}},\
  \bibinfo {pages} {053032} (\bibinfo {year} {2015})}\BibitemShut {NoStop}%
\bibitem [{\citenamefont {Aharon}\ \emph {et~al.}(2013)\citenamefont {Aharon},
  \citenamefont {Drewsen},\ and\ \citenamefont
  {Retzker}}]{aharon_general_2013}%
  \BibitemOpen
  \bibfield  {author} {\bibinfo {author} {\bibfnamefont {N.}~\bibnamefont
  {Aharon}}, \bibinfo {author} {\bibfnamefont {M.}~\bibnamefont {Drewsen}},\
  and\ \bibinfo {author} {\bibfnamefont {A.}~\bibnamefont {Retzker}},\ }\href
  {https://doi.org/10.1103/PhysRevLett.111.230507} {\bibfield  {journal}
  {\bibinfo  {journal} {Phys. Rev. Lett.}\ }\textbf {\bibinfo {volume} {111}},\
  \bibinfo {pages} {230507} (\bibinfo {year} {2013})}\BibitemShut {NoStop}%
\bibitem [{\citenamefont {Bermudez}\ \emph {et~al.}(2012)\citenamefont
  {Bermudez}, \citenamefont {Schmidt}, \citenamefont {Plenio},\ and\
  \citenamefont {Retzker}}]{bermudez_robust_2012}%
  \BibitemOpen
  \bibfield  {author} {\bibinfo {author} {\bibfnamefont {A.}~\bibnamefont
  {Bermudez}}, \bibinfo {author} {\bibfnamefont {P.~O.}\ \bibnamefont
  {Schmidt}}, \bibinfo {author} {\bibfnamefont {M.~B.}\ \bibnamefont
  {Plenio}},\ and\ \bibinfo {author} {\bibfnamefont {A.}~\bibnamefont
  {Retzker}},\ }\href {https://doi.org/10.1103/PhysRevA.85.040302} {\bibfield
  {journal} {\bibinfo  {journal} {Phys. Rev. A}\ }\textbf {\bibinfo {volume}
  {85}},\ \bibinfo {pages} {040302} (\bibinfo {year} {2012})}\BibitemShut
  {NoStop}%
\bibitem [{\citenamefont {Cai}\ \emph {et~al.}(2012)\citenamefont {Cai},
  \citenamefont {Naydenov}, \citenamefont {Pfeiffer}, \citenamefont
  {McGuinness}, \citenamefont {Jahnke}, \citenamefont {Jelezko}, \citenamefont
  {Plenio},\ and\ \citenamefont {Retzker}}]{cai_robust_2012}%
  \BibitemOpen
  \bibfield  {author} {\bibinfo {author} {\bibfnamefont {J.-M.}\ \bibnamefont
  {Cai}}, \bibinfo {author} {\bibfnamefont {B.}~\bibnamefont {Naydenov}},
  \bibinfo {author} {\bibfnamefont {R.}~\bibnamefont {Pfeiffer}}, \bibinfo
  {author} {\bibfnamefont {L.~P.}\ \bibnamefont {McGuinness}}, \bibinfo
  {author} {\bibfnamefont {K.~D.}\ \bibnamefont {Jahnke}}, \bibinfo {author}
  {\bibfnamefont {F.}~\bibnamefont {Jelezko}}, \bibinfo {author} {\bibfnamefont
  {M.~B.}\ \bibnamefont {Plenio}},\ and\ \bibinfo {author} {\bibfnamefont
  {A.}~\bibnamefont {Retzker}},\ }\href
  {https://doi.org/10.1088/1367-2630/14/11/113023} {\bibfield  {journal}
  {\bibinfo  {journal} {New J. Phys.}\ }\textbf {\bibinfo {volume} {14}},\
  \bibinfo {pages} {113023} (\bibinfo {year} {2012})}\BibitemShut {NoStop}%
\bibitem [{\citenamefont {Chaudhry}\ and\ \citenamefont
  {Gong}(2012)}]{chaudhry_decoherence_2012}%
  \BibitemOpen
  \bibfield  {author} {\bibinfo {author} {\bibfnamefont {A.~Z.}\ \bibnamefont
  {Chaudhry}}\ and\ \bibinfo {author} {\bibfnamefont {J.}~\bibnamefont
  {Gong}},\ }\href {https://doi.org/10.1103/PhysRevA.85.012315} {\bibfield
  {journal} {\bibinfo  {journal} {Phys. Rev. A}\ }\textbf {\bibinfo {volume}
  {85}},\ \bibinfo {pages} {012315} (\bibinfo {year} {2012})}\BibitemShut
  {NoStop}%
\bibitem [{\citenamefont {Facchi}\ \emph {et~al.}(2004)\citenamefont {Facchi},
  \citenamefont {Lidar},\ and\ \citenamefont
  {Pascazio}}]{facchi_unification_2004}%
  \BibitemOpen
  \bibfield  {author} {\bibinfo {author} {\bibfnamefont {P.}~\bibnamefont
  {Facchi}}, \bibinfo {author} {\bibfnamefont {D.~A.}\ \bibnamefont {Lidar}},\
  and\ \bibinfo {author} {\bibfnamefont {S.}~\bibnamefont {Pascazio}},\ }\href
  {https://doi.org/10.1103/PhysRevA.69.032314} {\bibfield  {journal} {\bibinfo
  {journal} {Phys. Rev. A}\ }\textbf {\bibinfo {volume} {69}},\ \bibinfo
  {pages} {032314} (\bibinfo {year} {2004})}\BibitemShut {NoStop}%
\bibitem [{\citenamefont {Viola}\ and\ \citenamefont
  {Lloyd}(1998)}]{viola_dynamical_1998}%
  \BibitemOpen
  \bibfield  {author} {\bibinfo {author} {\bibfnamefont {L.}~\bibnamefont
  {Viola}}\ and\ \bibinfo {author} {\bibfnamefont {S.}~\bibnamefont {Lloyd}},\
  }\href {https://doi.org/10.1103/PhysRevA.58.2733} {\bibfield  {journal}
  {\bibinfo  {journal} {Phys. Rev. A}\ }\textbf {\bibinfo {volume} {58}},\
  \bibinfo {pages} {2733} (\bibinfo {year} {1998})}\BibitemShut {NoStop}%
\bibitem [{\citenamefont {Morong}\ \emph {et~al.}(2023)\citenamefont {Morong},
  \citenamefont {Collins}, \citenamefont {De}, \citenamefont {Stavropoulos},
  \citenamefont {You},\ and\ \citenamefont {Monroe}}]{morong_engineering_2023}%
  \BibitemOpen
  \bibfield  {author} {\bibinfo {author} {\bibfnamefont {W.}~\bibnamefont
  {Morong}}, \bibinfo {author} {\bibfnamefont {K.}~\bibnamefont {Collins}},
  \bibinfo {author} {\bibfnamefont {A.}~\bibnamefont {De}}, \bibinfo {author}
  {\bibfnamefont {E.}~\bibnamefont {Stavropoulos}}, \bibinfo {author}
  {\bibfnamefont {T.}~\bibnamefont {You}},\ and\ \bibinfo {author}
  {\bibfnamefont {C.}~\bibnamefont {Monroe}},\ }\href
  {https://doi.org/10.1103/PRXQuantum.4.010334} {\bibfield  {journal} {\bibinfo
   {journal} {PRX Quantum}\ }\textbf {\bibinfo {volume} {4}},\ \bibinfo {pages}
  {010334} (\bibinfo {year} {2023})}\BibitemShut {NoStop}%
\bibitem [{\citenamefont {Barthel}\ \emph {et~al.}(2023)\citenamefont
  {Barthel}, \citenamefont {Huber}, \citenamefont {Casanova}, \citenamefont
  {Arrazola}, \citenamefont {Niroomand}, \citenamefont {Sriarunothai},
  \citenamefont {Plenio},\ and\ \citenamefont
  {Wunderlich}}]{barthel_robust_2023}%
  \BibitemOpen
  \bibfield  {author} {\bibinfo {author} {\bibfnamefont {P.}~\bibnamefont
  {Barthel}}, \bibinfo {author} {\bibfnamefont {P.~H.}\ \bibnamefont {Huber}},
  \bibinfo {author} {\bibfnamefont {J.}~\bibnamefont {Casanova}}, \bibinfo
  {author} {\bibfnamefont {I.}~\bibnamefont {Arrazola}}, \bibinfo {author}
  {\bibfnamefont {D.}~\bibnamefont {Niroomand}}, \bibinfo {author}
  {\bibfnamefont {T.}~\bibnamefont {Sriarunothai}}, \bibinfo {author}
  {\bibfnamefont {M.~B.}\ \bibnamefont {Plenio}},\ and\ \bibinfo {author}
  {\bibfnamefont {C.}~\bibnamefont {Wunderlich}},\ }\href
  {https://doi.org/10.1088/1367-2630/acd4db} {\bibfield  {journal} {\bibinfo
  {journal} {New J. Phys.}\ }\textbf {\bibinfo {volume} {25}},\ \bibinfo
  {pages} {063023} (\bibinfo {year} {2023})}\BibitemShut {NoStop}%
\bibitem [{\citenamefont {Leu}\ \emph {et~al.}(2023)\citenamefont {Leu},
  \citenamefont {Gely}, \citenamefont {Weber}, \citenamefont {Smith},
  \citenamefont {Nadlinger},\ and\ \citenamefont {Lucas}}]{leu_fast_2023-1}%
  \BibitemOpen
  \bibfield  {author} {\bibinfo {author} {\bibfnamefont {A.~D.}\ \bibnamefont
  {Leu}}, \bibinfo {author} {\bibfnamefont {M.~F.}\ \bibnamefont {Gely}},
  \bibinfo {author} {\bibfnamefont {M.~A.}\ \bibnamefont {Weber}}, \bibinfo
  {author} {\bibfnamefont {M.~C.}\ \bibnamefont {Smith}}, \bibinfo {author}
  {\bibfnamefont {D.~P.}\ \bibnamefont {Nadlinger}},\ and\ \bibinfo {author}
  {\bibfnamefont {D.~M.}\ \bibnamefont {Lucas}},\ }\href
  {https://doi.org/10.1103/PhysRevLett.131.120601} {\bibfield  {journal}
  {\bibinfo  {journal} {Phys. Rev. Lett.}\ }\textbf {\bibinfo {volume} {131}},\
  \bibinfo {pages} {120601} (\bibinfo {year} {2023})}\BibitemShut {NoStop}%
\bibitem [{\citenamefont {Fang}\ \emph {et~al.}(2022)\citenamefont {Fang},
  \citenamefont {Wang}, \citenamefont {Huang}, \citenamefont {Brown},\ and\
  \citenamefont {Kim}}]{fang_crosstalk_2022}%
  \BibitemOpen
  \bibfield  {author} {\bibinfo {author} {\bibfnamefont {C.}~\bibnamefont
  {Fang}}, \bibinfo {author} {\bibfnamefont {Y.}~\bibnamefont {Wang}}, \bibinfo
  {author} {\bibfnamefont {S.}~\bibnamefont {Huang}}, \bibinfo {author}
  {\bibfnamefont {K.~R.}\ \bibnamefont {Brown}},\ and\ \bibinfo {author}
  {\bibfnamefont {J.}~\bibnamefont {Kim}},\ }\href
  {https://doi.org/10.1103/PhysRevLett.129.240504} {\bibfield  {journal}
  {\bibinfo  {journal} {Phys. Rev. Lett.}\ }\textbf {\bibinfo {volume} {129}},\
  \bibinfo {pages} {240504} (\bibinfo {year} {2022})}\BibitemShut {NoStop}%
\bibitem [{\citenamefont {Pokharel}\ \emph {et~al.}(2018)\citenamefont
  {Pokharel}, \citenamefont {Anand}, \citenamefont {Fortman},\ and\
  \citenamefont {Lidar}}]{pokharel_demonstration_2018}%
  \BibitemOpen
  \bibfield  {author} {\bibinfo {author} {\bibfnamefont {B.}~\bibnamefont
  {Pokharel}}, \bibinfo {author} {\bibfnamefont {N.}~\bibnamefont {Anand}},
  \bibinfo {author} {\bibfnamefont {B.}~\bibnamefont {Fortman}},\ and\ \bibinfo
  {author} {\bibfnamefont {D.~A.}\ \bibnamefont {Lidar}},\ }\href
  {https://doi.org/10.1103/PhysRevLett.121.220502} {\bibfield  {journal}
  {\bibinfo  {journal} {Phys. Rev. Lett.}\ }\textbf {\bibinfo {volume} {121}},\
  \bibinfo {pages} {220502} (\bibinfo {year} {2018})}\BibitemShut {NoStop}%
\bibitem [{\citenamefont {Stark}\ \emph {et~al.}(2018)\citenamefont {Stark},
  \citenamefont {Aharon}, \citenamefont {Huck}, \citenamefont {{El-Ella}},
  \citenamefont {Retzker}, \citenamefont {Jelezko},\ and\ \citenamefont
  {Andersen}}]{stark_clock_2018}%
  \BibitemOpen
  \bibfield  {author} {\bibinfo {author} {\bibfnamefont {A.}~\bibnamefont
  {Stark}}, \bibinfo {author} {\bibfnamefont {N.}~\bibnamefont {Aharon}},
  \bibinfo {author} {\bibfnamefont {A.}~\bibnamefont {Huck}}, \bibinfo {author}
  {\bibfnamefont {H.~A.~R.}\ \bibnamefont {{El-Ella}}}, \bibinfo {author}
  {\bibfnamefont {A.}~\bibnamefont {Retzker}}, \bibinfo {author} {\bibfnamefont
  {F.}~\bibnamefont {Jelezko}},\ and\ \bibinfo {author} {\bibfnamefont {U.~L.}\
  \bibnamefont {Andersen}},\ }\href
  {https://doi.org/10.1038/s41598-018-31984-4} {\bibfield  {journal} {\bibinfo
  {journal} {Sci. Rep.}\ }\textbf {\bibinfo {volume} {8}},\ \bibinfo {pages}
  {14807} (\bibinfo {year} {2018})}\BibitemShut {NoStop}%
\bibitem [{\citenamefont {Randall}\ \emph {et~al.}(2015)\citenamefont
  {Randall}, \citenamefont {Weidt}, \citenamefont {Standing}, \citenamefont
  {Lake}, \citenamefont {Webster}, \citenamefont {Murgia}, \citenamefont
  {Navickas}, \citenamefont {Roth},\ and\ \citenamefont
  {Hensinger}}]{randall_efficient_2015}%
  \BibitemOpen
  \bibfield  {author} {\bibinfo {author} {\bibfnamefont {J.}~\bibnamefont
  {Randall}}, \bibinfo {author} {\bibfnamefont {S.}~\bibnamefont {Weidt}},
  \bibinfo {author} {\bibfnamefont {E.~D.}\ \bibnamefont {Standing}}, \bibinfo
  {author} {\bibfnamefont {K.}~\bibnamefont {Lake}}, \bibinfo {author}
  {\bibfnamefont {S.~C.}\ \bibnamefont {Webster}}, \bibinfo {author}
  {\bibfnamefont {D.~F.}\ \bibnamefont {Murgia}}, \bibinfo {author}
  {\bibfnamefont {T.}~\bibnamefont {Navickas}}, \bibinfo {author}
  {\bibfnamefont {K.}~\bibnamefont {Roth}},\ and\ \bibinfo {author}
  {\bibfnamefont {W.~K.}\ \bibnamefont {Hensinger}},\ }\href
  {https://doi.org/10.1103/PhysRevA.91.012322} {\bibfield  {journal} {\bibinfo
  {journal} {Phys. Rev. A}\ }\textbf {\bibinfo {volume} {91}},\ \bibinfo
  {pages} {012322} (\bibinfo {year} {2015})}\BibitemShut {NoStop}%
\bibitem [{\citenamefont {Tan}\ \emph {et~al.}(2013)\citenamefont {Tan},
  \citenamefont {Gaebler}, \citenamefont {Bowler}, \citenamefont {Lin},
  \citenamefont {Jost}, \citenamefont {Leibfried},\ and\ \citenamefont
  {Wineland}}]{tan_demonstration_2013}%
  \BibitemOpen
  \bibfield  {author} {\bibinfo {author} {\bibfnamefont {T.~R.}\ \bibnamefont
  {Tan}}, \bibinfo {author} {\bibfnamefont {J.~P.}\ \bibnamefont {Gaebler}},
  \bibinfo {author} {\bibfnamefont {R.}~\bibnamefont {Bowler}}, \bibinfo
  {author} {\bibfnamefont {Y.}~\bibnamefont {Lin}}, \bibinfo {author}
  {\bibfnamefont {J.~D.}\ \bibnamefont {Jost}}, \bibinfo {author}
  {\bibfnamefont {D.}~\bibnamefont {Leibfried}},\ and\ \bibinfo {author}
  {\bibfnamefont {D.~J.}\ \bibnamefont {Wineland}},\ }\href
  {https://doi.org/10.1103/PhysRevLett.110.263002} {\bibfield  {journal}
  {\bibinfo  {journal} {Phys. Rev. Lett.}\ }\textbf {\bibinfo {volume} {110}},\
  \bibinfo {pages} {263002} (\bibinfo {year} {2013})}\BibitemShut {NoStop}%
\bibitem [{\citenamefont {Webster}\ \emph {et~al.}(2013)\citenamefont
  {Webster}, \citenamefont {Weidt}, \citenamefont {Lake}, \citenamefont
  {McLoughlin},\ and\ \citenamefont {Hensinger}}]{webster_simple_2013}%
  \BibitemOpen
  \bibfield  {author} {\bibinfo {author} {\bibfnamefont {S.~C.}\ \bibnamefont
  {Webster}}, \bibinfo {author} {\bibfnamefont {S.}~\bibnamefont {Weidt}},
  \bibinfo {author} {\bibfnamefont {K.}~\bibnamefont {Lake}}, \bibinfo {author}
  {\bibfnamefont {J.~J.}\ \bibnamefont {McLoughlin}},\ and\ \bibinfo {author}
  {\bibfnamefont {W.~K.}\ \bibnamefont {Hensinger}},\ }\href
  {https://doi.org/10.1103/PhysRevLett.111.140501} {\bibfield  {journal}
  {\bibinfo  {journal} {Phys. Rev. Lett.}\ }\textbf {\bibinfo {volume} {111}},\
  \bibinfo {pages} {140501} (\bibinfo {year} {2013})}\BibitemShut {NoStop}%
\bibitem [{\citenamefont {Souza}\ \emph {et~al.}(2012)\citenamefont {Souza},
  \citenamefont {{\'A}lvarez},\ and\ \citenamefont
  {Suter}}]{souza_experimental_2012}%
  \BibitemOpen
  \bibfield  {author} {\bibinfo {author} {\bibfnamefont {A.~M.}\ \bibnamefont
  {Souza}}, \bibinfo {author} {\bibfnamefont {G.~A.}\ \bibnamefont
  {{\'A}lvarez}},\ and\ \bibinfo {author} {\bibfnamefont {D.}~\bibnamefont
  {Suter}},\ }\href {https://doi.org/10.1103/PhysRevA.86.050301} {\bibfield
  {journal} {\bibinfo  {journal} {Phys. Rev. A}\ }\textbf {\bibinfo {volume}
  {86}},\ \bibinfo {pages} {050301} (\bibinfo {year} {2012})}\BibitemShut
  {NoStop}%
\bibitem [{\citenamefont {Noguchi}\ \emph {et~al.}(2012)\citenamefont
  {Noguchi}, \citenamefont {Haze}, \citenamefont {Toyoda},\ and\ \citenamefont
  {Urabe}}]{noguchi_generation_2012}%
  \BibitemOpen
  \bibfield  {author} {\bibinfo {author} {\bibfnamefont {A.}~\bibnamefont
  {Noguchi}}, \bibinfo {author} {\bibfnamefont {S.}~\bibnamefont {Haze}},
  \bibinfo {author} {\bibfnamefont {K.}~\bibnamefont {Toyoda}},\ and\ \bibinfo
  {author} {\bibfnamefont {S.}~\bibnamefont {Urabe}},\ }\href
  {https://doi.org/10.1103/PhysRevLett.108.060503} {\bibfield  {journal}
  {\bibinfo  {journal} {Phys. Rev. Lett.}\ }\textbf {\bibinfo {volume} {108}},\
  \bibinfo {pages} {060503} (\bibinfo {year} {2012})}\BibitemShut {NoStop}%
\bibitem [{\citenamefont {Xu}\ \emph {et~al.}(2012)\citenamefont {Xu},
  \citenamefont {Wang}, \citenamefont {Duan}, \citenamefont {Huang},
  \citenamefont {Wang}, \citenamefont {Wang}, \citenamefont {Xu}, \citenamefont
  {Kong}, \citenamefont {Shi}, \citenamefont {Rong},\ and\ \citenamefont
  {Du}}]{xu_coherence-protected_2012}%
  \BibitemOpen
  \bibfield  {author} {\bibinfo {author} {\bibfnamefont {X.}~\bibnamefont
  {Xu}}, \bibinfo {author} {\bibfnamefont {Z.}~\bibnamefont {Wang}}, \bibinfo
  {author} {\bibfnamefont {C.}~\bibnamefont {Duan}}, \bibinfo {author}
  {\bibfnamefont {P.}~\bibnamefont {Huang}}, \bibinfo {author} {\bibfnamefont
  {P.}~\bibnamefont {Wang}}, \bibinfo {author} {\bibfnamefont {Y.}~\bibnamefont
  {Wang}}, \bibinfo {author} {\bibfnamefont {N.}~\bibnamefont {Xu}}, \bibinfo
  {author} {\bibfnamefont {X.}~\bibnamefont {Kong}}, \bibinfo {author}
  {\bibfnamefont {F.}~\bibnamefont {Shi}}, \bibinfo {author} {\bibfnamefont
  {X.}~\bibnamefont {Rong}},\ and\ \bibinfo {author} {\bibfnamefont
  {J.}~\bibnamefont {Du}},\ }\href
  {https://doi.org/10.1103/PhysRevLett.109.070502} {\bibfield  {journal}
  {\bibinfo  {journal} {Phys. Rev. Lett.}\ }\textbf {\bibinfo {volume} {109}},\
  \bibinfo {pages} {070502} (\bibinfo {year} {2012})}\BibitemShut {NoStop}%
\bibitem [{\citenamefont {Zalivako}\ \emph {et~al.}(2023)\citenamefont
  {Zalivako}, \citenamefont {Borisenko}, \citenamefont {Semerikov},
  \citenamefont {Korolkov}, \citenamefont {Sidorov}, \citenamefont {Galstyan},
  \citenamefont {Semenin}, \citenamefont {Smirnov}, \citenamefont {Aksenov},
  \citenamefont {Fedorov}, \citenamefont {Khabarova},\ and\ \citenamefont
  {Kolachevsky}}]{zalivako_continuous_2023}%
  \BibitemOpen
  \bibfield  {author} {\bibinfo {author} {\bibfnamefont {I.~V.}\ \bibnamefont
  {Zalivako}}, \bibinfo {author} {\bibfnamefont {A.~S.}\ \bibnamefont
  {Borisenko}}, \bibinfo {author} {\bibfnamefont {I.~A.}\ \bibnamefont
  {Semerikov}}, \bibinfo {author} {\bibfnamefont {A.}~\bibnamefont {Korolkov}},
  \bibinfo {author} {\bibfnamefont {P.~L.}\ \bibnamefont {Sidorov}}, \bibinfo
  {author} {\bibfnamefont {K.}~\bibnamefont {Galstyan}}, \bibinfo {author}
  {\bibfnamefont {N.~V.}\ \bibnamefont {Semenin}}, \bibinfo {author}
  {\bibfnamefont {V.}~\bibnamefont {Smirnov}}, \bibinfo {author} {\bibfnamefont
  {M.~A.}\ \bibnamefont {Aksenov}}, \bibinfo {author} {\bibfnamefont {A.~K.}\
  \bibnamefont {Fedorov}}, \bibinfo {author} {\bibfnamefont {K.~Y.}\
  \bibnamefont {Khabarova}},\ and\ \bibinfo {author} {\bibfnamefont {N.~N.}\
  \bibnamefont {Kolachevsky}},\ }\href
  {https://doi.org/10.3389/frqst.2023.1228208} {\bibfield  {journal} {\bibinfo
  {journal} {Frontiers in Quantum Science and Technology}\ }\textbf {\bibinfo
  {volume} {2}},\ \bibinfo {pages} {1228208} (\bibinfo {year} {2023})},\
  \Eprint {https://arxiv.org/abs/2305.06071} {arxiv:2305.06071 [quant-ph]}
  \BibitemShut {NoStop}%
\bibitem [{\citenamefont {Wang}\ \emph {et~al.}(2021)\citenamefont {Wang},
  \citenamefont {Luan}, \citenamefont {Qiao}, \citenamefont {Um}, \citenamefont
  {Zhang}, \citenamefont {Wang}, \citenamefont {Yuan}, \citenamefont {Gu},
  \citenamefont {Zhang},\ and\ \citenamefont {Kim}}]{wang_single_2021}%
  \BibitemOpen
  \bibfield  {author} {\bibinfo {author} {\bibfnamefont {P.}~\bibnamefont
  {Wang}}, \bibinfo {author} {\bibfnamefont {C.-Y.}\ \bibnamefont {Luan}},
  \bibinfo {author} {\bibfnamefont {M.}~\bibnamefont {Qiao}}, \bibinfo {author}
  {\bibfnamefont {M.}~\bibnamefont {Um}}, \bibinfo {author} {\bibfnamefont
  {J.}~\bibnamefont {Zhang}}, \bibinfo {author} {\bibfnamefont
  {Y.}~\bibnamefont {Wang}}, \bibinfo {author} {\bibfnamefont {X.}~\bibnamefont
  {Yuan}}, \bibinfo {author} {\bibfnamefont {M.}~\bibnamefont {Gu}}, \bibinfo
  {author} {\bibfnamefont {J.}~\bibnamefont {Zhang}},\ and\ \bibinfo {author}
  {\bibfnamefont {K.}~\bibnamefont {Kim}},\ }\href
  {https://doi.org/10.1038/s41467-020-20330-w} {\bibfield  {journal} {\bibinfo
  {journal} {Nat. Commun.}\ }\textbf {\bibinfo {volume} {12}},\ \bibinfo
  {pages} {233} (\bibinfo {year} {2021})}\BibitemShut {NoStop}%
\bibitem [{\citenamefont {{Sinuco-Leon}}\ \emph {et~al.}(2021)\citenamefont
  {{Sinuco-Leon}}, \citenamefont {Mas}, \citenamefont {Pandey}, \citenamefont
  {Vasilakis}, \citenamefont {Garraway},\ and\ \citenamefont {{von
  Klitzing}}}]{sinuco-leon_decoherence-free_2021}%
  \BibitemOpen
  \bibfield  {author} {\bibinfo {author} {\bibfnamefont {G.~A.}\ \bibnamefont
  {{Sinuco-Leon}}}, \bibinfo {author} {\bibfnamefont {H.}~\bibnamefont {Mas}},
  \bibinfo {author} {\bibfnamefont {S.}~\bibnamefont {Pandey}}, \bibinfo
  {author} {\bibfnamefont {G.}~\bibnamefont {Vasilakis}}, \bibinfo {author}
  {\bibfnamefont {B.~M.}\ \bibnamefont {Garraway}},\ and\ \bibinfo {author}
  {\bibfnamefont {W.}~\bibnamefont {{von Klitzing}}},\ }\href
  {https://doi.org/10.1103/PhysRevA.104.033307} {\bibfield  {journal} {\bibinfo
   {journal} {Phys. Rev. A}\ }\textbf {\bibinfo {volume} {104}},\ \bibinfo
  {pages} {033307} (\bibinfo {year} {2021})}\BibitemShut {NoStop}%
\bibitem [{\citenamefont {{Sinuco-Leon}}\ \emph {et~al.}(2019)\citenamefont
  {{Sinuco-Leon}}, \citenamefont {Garraway}, \citenamefont {Mas}, \citenamefont
  {Pandey}, \citenamefont {Vasilakis}, \citenamefont {Bolpasi}, \citenamefont
  {{von Klitzing}}, \citenamefont {Foxon}, \citenamefont {Jammi}, \citenamefont
  {Poulios},\ and\ \citenamefont {Fernholz}}]{sinuco-leon_microwave_2019}%
  \BibitemOpen
  \bibfield  {author} {\bibinfo {author} {\bibfnamefont {G.~A.}\ \bibnamefont
  {{Sinuco-Leon}}}, \bibinfo {author} {\bibfnamefont {B.~M.}\ \bibnamefont
  {Garraway}}, \bibinfo {author} {\bibfnamefont {H.}~\bibnamefont {Mas}},
  \bibinfo {author} {\bibfnamefont {S.}~\bibnamefont {Pandey}}, \bibinfo
  {author} {\bibfnamefont {G.}~\bibnamefont {Vasilakis}}, \bibinfo {author}
  {\bibfnamefont {V.}~\bibnamefont {Bolpasi}}, \bibinfo {author} {\bibfnamefont
  {W.}~\bibnamefont {{von Klitzing}}}, \bibinfo {author} {\bibfnamefont
  {B.}~\bibnamefont {Foxon}}, \bibinfo {author} {\bibfnamefont
  {S.}~\bibnamefont {Jammi}}, \bibinfo {author} {\bibfnamefont
  {K.}~\bibnamefont {Poulios}},\ and\ \bibinfo {author} {\bibfnamefont
  {T.}~\bibnamefont {Fernholz}},\ }\href
  {https://doi.org/10.1103/PhysRevA.100.053416} {\bibfield  {journal} {\bibinfo
   {journal} {Phys. Rev. A}\ }\textbf {\bibinfo {volume} {100}},\ \bibinfo
  {pages} {053416} (\bibinfo {year} {2019})}\BibitemShut {NoStop}%
\bibitem [{\citenamefont {Trypogeorgos}\ \emph {et~al.}(2018)\citenamefont
  {Trypogeorgos}, \citenamefont {{Vald{\'e}s-Curiel}}, \citenamefont
  {Lundblad},\ and\ \citenamefont {Spielman}}]{trypogeorgos_synthetic_2018}%
  \BibitemOpen
  \bibfield  {author} {\bibinfo {author} {\bibfnamefont {D.}~\bibnamefont
  {Trypogeorgos}}, \bibinfo {author} {\bibfnamefont {A.}~\bibnamefont
  {{Vald{\'e}s-Curiel}}}, \bibinfo {author} {\bibfnamefont {N.}~\bibnamefont
  {Lundblad}},\ and\ \bibinfo {author} {\bibfnamefont {I.~B.}\ \bibnamefont
  {Spielman}},\ }\href {https://doi.org/10.1103/PhysRevA.97.013407} {\bibfield
  {journal} {\bibinfo  {journal} {Phys. Rev. A}\ }\textbf {\bibinfo {volume}
  {97}},\ \bibinfo {pages} {013407} (\bibinfo {year} {2018})}\BibitemShut
  {NoStop}%
\bibitem [{\citenamefont {Laucht}\ \emph {et~al.}(2017)\citenamefont {Laucht},
  \citenamefont {Kalra}, \citenamefont {Simmons}, \citenamefont {Dehollain},
  \citenamefont {Muhonen}, \citenamefont {Mohiyaddin}, \citenamefont {Freer},
  \citenamefont {Hudson}, \citenamefont {Itoh}, \citenamefont {Jamieson},
  \citenamefont {McCallum}, \citenamefont {Dzurak},\ and\ \citenamefont
  {Morello}}]{laucht_dressed_2017}%
  \BibitemOpen
  \bibfield  {author} {\bibinfo {author} {\bibfnamefont {A.}~\bibnamefont
  {Laucht}}, \bibinfo {author} {\bibfnamefont {R.}~\bibnamefont {Kalra}},
  \bibinfo {author} {\bibfnamefont {S.}~\bibnamefont {Simmons}}, \bibinfo
  {author} {\bibfnamefont {J.~P.}\ \bibnamefont {Dehollain}}, \bibinfo {author}
  {\bibfnamefont {J.~T.}\ \bibnamefont {Muhonen}}, \bibinfo {author}
  {\bibfnamefont {F.~A.}\ \bibnamefont {Mohiyaddin}}, \bibinfo {author}
  {\bibfnamefont {S.}~\bibnamefont {Freer}}, \bibinfo {author} {\bibfnamefont
  {F.~E.}\ \bibnamefont {Hudson}}, \bibinfo {author} {\bibfnamefont {K.~M.}\
  \bibnamefont {Itoh}}, \bibinfo {author} {\bibfnamefont {D.~N.}\ \bibnamefont
  {Jamieson}}, \bibinfo {author} {\bibfnamefont {J.~C.}\ \bibnamefont
  {McCallum}}, \bibinfo {author} {\bibfnamefont {A.~S.}\ \bibnamefont
  {Dzurak}},\ and\ \bibinfo {author} {\bibfnamefont {A.}~\bibnamefont
  {Morello}},\ }\href {https://doi.org/10.1038/nnano.2016.178} {\bibfield
  {journal} {\bibinfo  {journal} {Nat. Nanotechnol.}\ }\textbf {\bibinfo
  {volume} {12}},\ \bibinfo {pages} {61} (\bibinfo {year} {2017})}\BibitemShut
  {NoStop}%
\bibitem [{\citenamefont {Zhang}\ \emph {et~al.}(2014)\citenamefont {Zhang},
  \citenamefont {Souza}, \citenamefont {Brandao},\ and\ \citenamefont
  {Suter}}]{zhang_protected_2014}%
  \BibitemOpen
  \bibfield  {author} {\bibinfo {author} {\bibfnamefont {J.}~\bibnamefont
  {Zhang}}, \bibinfo {author} {\bibfnamefont {A.~M.}\ \bibnamefont {Souza}},
  \bibinfo {author} {\bibfnamefont {F.~D.}\ \bibnamefont {Brandao}},\ and\
  \bibinfo {author} {\bibfnamefont {D.}~\bibnamefont {Suter}},\ }\href
  {https://doi.org/10.1103/PhysRevLett.112.050502} {\bibfield  {journal}
  {\bibinfo  {journal} {Phys. Rev. Lett.}\ }\textbf {\bibinfo {volume} {112}},\
  \bibinfo {pages} {050502} (\bibinfo {year} {2014})}\BibitemShut {NoStop}%
\bibitem [{\citenamefont {Golter}\ \emph {et~al.}(2014)\citenamefont {Golter},
  \citenamefont {Baldwin},\ and\ \citenamefont
  {Wang}}]{golter_protecting_2014}%
  \BibitemOpen
  \bibfield  {author} {\bibinfo {author} {\bibfnamefont {D.~A.}\ \bibnamefont
  {Golter}}, \bibinfo {author} {\bibfnamefont {T.~K.}\ \bibnamefont
  {Baldwin}},\ and\ \bibinfo {author} {\bibfnamefont {H.}~\bibnamefont
  {Wang}},\ }\href {https://doi.org/10.1103/PhysRevLett.113.237601} {\bibfield
  {journal} {\bibinfo  {journal} {Phys. Rev. Lett.}\ }\textbf {\bibinfo
  {volume} {113}},\ \bibinfo {pages} {237601} (\bibinfo {year}
  {2014})}\BibitemShut {NoStop}%
\bibitem [{\citenamefont {Liu}\ \emph {et~al.}(2013)\citenamefont {Liu},
  \citenamefont {Po}, \citenamefont {Du}, \citenamefont {Liu},\ and\
  \citenamefont {Pan}}]{liu_noise-resilient_2013}%
  \BibitemOpen
  \bibfield  {author} {\bibinfo {author} {\bibfnamefont {G.-Q.}\ \bibnamefont
  {Liu}}, \bibinfo {author} {\bibfnamefont {H.~C.}\ \bibnamefont {Po}},
  \bibinfo {author} {\bibfnamefont {J.}~\bibnamefont {Du}}, \bibinfo {author}
  {\bibfnamefont {R.-B.}\ \bibnamefont {Liu}},\ and\ \bibinfo {author}
  {\bibfnamefont {X.-Y.}\ \bibnamefont {Pan}},\ }\href
  {https://doi.org/10.1038/ncomms3254} {\bibfield  {journal} {\bibinfo
  {journal} {Nat. Commun.}\ }\textbf {\bibinfo {volume} {4}},\ \bibinfo {pages}
  {2254} (\bibinfo {year} {2013})}\BibitemShut {NoStop}%
\bibitem [{\citenamefont {Biercuk}\ \emph {et~al.}(2009)\citenamefont
  {Biercuk}, \citenamefont {Uys}, \citenamefont {VanDevender}, \citenamefont
  {Shiga}, \citenamefont {Itano},\ and\ \citenamefont
  {Bollinger}}]{biercuk_optimized_2009}%
  \BibitemOpen
  \bibfield  {author} {\bibinfo {author} {\bibfnamefont {M.~J.}\ \bibnamefont
  {Biercuk}}, \bibinfo {author} {\bibfnamefont {H.}~\bibnamefont {Uys}},
  \bibinfo {author} {\bibfnamefont {A.~P.}\ \bibnamefont {VanDevender}},
  \bibinfo {author} {\bibfnamefont {N.}~\bibnamefont {Shiga}}, \bibinfo
  {author} {\bibfnamefont {W.~M.}\ \bibnamefont {Itano}},\ and\ \bibinfo
  {author} {\bibfnamefont {J.~J.}\ \bibnamefont {Bollinger}},\ }\href
  {https://doi.org/10.1038/nature07951} {\bibfield  {journal} {\bibinfo
  {journal} {Nature}\ }\textbf {\bibinfo {volume} {458}},\ \bibinfo {pages}
  {996} (\bibinfo {year} {2009})}\BibitemShut {NoStop}%
\bibitem [{\citenamefont {Timoney}\ \emph {et~al.}(2011)\citenamefont
  {Timoney}, \citenamefont {Baumgart}, \citenamefont {Johanning}, \citenamefont
  {Varon}, \citenamefont {Plenio}, \citenamefont {Retzker},\ and\ \citenamefont
  {Wunderlich}}]{timoney_quantum_2011}%
  \BibitemOpen
  \bibfield  {author} {\bibinfo {author} {\bibfnamefont {N.}~\bibnamefont
  {Timoney}}, \bibinfo {author} {\bibfnamefont {I.}~\bibnamefont {Baumgart}},
  \bibinfo {author} {\bibfnamefont {M.}~\bibnamefont {Johanning}}, \bibinfo
  {author} {\bibfnamefont {A.~F.}\ \bibnamefont {Varon}}, \bibinfo {author}
  {\bibfnamefont {M.~B.}\ \bibnamefont {Plenio}}, \bibinfo {author}
  {\bibfnamefont {A.}~\bibnamefont {Retzker}},\ and\ \bibinfo {author}
  {\bibfnamefont {{\relax Ch}.}~\bibnamefont {Wunderlich}},\ }\href
  {https://doi.org/10.1038/nature10319} {\bibfield  {journal} {\bibinfo
  {journal} {Nature}\ }\textbf {\bibinfo {volume} {476}},\ \bibinfo {pages}
  {185} (\bibinfo {year} {2011})}\BibitemShut {NoStop}%
\bibitem [{\citenamefont {de~Lange}\ \emph {et~al.}(2010)\citenamefont
  {de~Lange}, \citenamefont {Wang}, \citenamefont {Rist{\`e}}, \citenamefont
  {Dobrovitski},\ and\ \citenamefont {Hanson}}]{lange_universal_2010}%
  \BibitemOpen
  \bibfield  {author} {\bibinfo {author} {\bibfnamefont {G.}~\bibnamefont
  {de~Lange}}, \bibinfo {author} {\bibfnamefont {Z.~H.}\ \bibnamefont {Wang}},
  \bibinfo {author} {\bibfnamefont {D.}~\bibnamefont {Rist{\`e}}}, \bibinfo
  {author} {\bibfnamefont {V.~V.}\ \bibnamefont {Dobrovitski}},\ and\ \bibinfo
  {author} {\bibfnamefont {R.}~\bibnamefont {Hanson}},\ }\href
  {https://doi.org/10.1126/science.1192739} {\bibfield  {journal} {\bibinfo
  {journal} {Science}\ }\textbf {\bibinfo {volume} {330}},\ \bibinfo {pages}
  {60} (\bibinfo {year} {2010})}\BibitemShut {NoStop}%
\bibitem [{\citenamefont {Zhang}\ \emph {et~al.}(2021)\citenamefont {Zhang},
  \citenamefont {Xie}, \citenamefont {Zhang}, \citenamefont {Wang},
  \citenamefont {Wu}, \citenamefont {Chen}, \citenamefont {Wu},\ and\
  \citenamefont {Chen}}]{zhang_estimation_2021}%
  \BibitemOpen
  \bibfield  {author} {\bibinfo {author} {\bibfnamefont {M.}~\bibnamefont
  {Zhang}}, \bibinfo {author} {\bibfnamefont {Y.}~\bibnamefont {Xie}}, \bibinfo
  {author} {\bibfnamefont {J.}~\bibnamefont {Zhang}}, \bibinfo {author}
  {\bibfnamefont {W.}~\bibnamefont {Wang}}, \bibinfo {author} {\bibfnamefont
  {C.}~\bibnamefont {Wu}}, \bibinfo {author} {\bibfnamefont {T.}~\bibnamefont
  {Chen}}, \bibinfo {author} {\bibfnamefont {W.}~\bibnamefont {Wu}},\ and\
  \bibinfo {author} {\bibfnamefont {P.}~\bibnamefont {Chen}},\ }\href
  {https://doi.org/10.1103/PhysRevApplied.15.014033} {\bibfield  {journal}
  {\bibinfo  {journal} {Phys. Rev. Applied}\ }\textbf {\bibinfo {volume}
  {15}},\ \bibinfo {pages} {014033} (\bibinfo {year} {2021})}\BibitemShut
  {NoStop}%
\bibitem [{\citenamefont {Baumgart}\ \emph {et~al.}(2016)\citenamefont
  {Baumgart}, \citenamefont {Cai}, \citenamefont {Retzker}, \citenamefont
  {Plenio},\ and\ \citenamefont {Wunderlich}}]{baumgart_ultrasensitive_2016}%
  \BibitemOpen
  \bibfield  {author} {\bibinfo {author} {\bibfnamefont {I.}~\bibnamefont
  {Baumgart}}, \bibinfo {author} {\bibfnamefont {J.-M.}\ \bibnamefont {Cai}},
  \bibinfo {author} {\bibfnamefont {A.}~\bibnamefont {Retzker}}, \bibinfo
  {author} {\bibfnamefont {M.~B.}\ \bibnamefont {Plenio}},\ and\ \bibinfo
  {author} {\bibfnamefont {{\relax Ch}.}~\bibnamefont {Wunderlich}},\ }\href
  {https://doi.org/10.1103/PhysRevLett.116.240801} {\bibfield  {journal}
  {\bibinfo  {journal} {Phys. Rev. Lett.}\ }\textbf {\bibinfo {volume} {116}},\
  \bibinfo {pages} {240801} (\bibinfo {year} {2016})}\BibitemShut {NoStop}%
\bibitem [{\citenamefont {Kotler}\ \emph {et~al.}(2011)\citenamefont {Kotler},
  \citenamefont {Akerman}, \citenamefont {Glickman}, \citenamefont {Keselman},\
  and\ \citenamefont {Ozeri}}]{kotler_single-ion_2011}%
  \BibitemOpen
  \bibfield  {author} {\bibinfo {author} {\bibfnamefont {S.}~\bibnamefont
  {Kotler}}, \bibinfo {author} {\bibfnamefont {N.}~\bibnamefont {Akerman}},
  \bibinfo {author} {\bibfnamefont {Y.}~\bibnamefont {Glickman}}, \bibinfo
  {author} {\bibfnamefont {A.}~\bibnamefont {Keselman}},\ and\ \bibinfo
  {author} {\bibfnamefont {R.}~\bibnamefont {Ozeri}},\ }\href
  {https://doi.org/10.1038/nature10010} {\bibfield  {journal} {\bibinfo
  {journal} {Nature}\ }\textbf {\bibinfo {volume} {473}},\ \bibinfo {pages}
  {61} (\bibinfo {year} {2011})}\BibitemShut {NoStop}%
\bibitem [{\citenamefont {Hall}\ \emph {et~al.}(2010)\citenamefont {Hall},
  \citenamefont {Hill}, \citenamefont {Cole},\ and\ \citenamefont
  {Hollenberg}}]{hall_ultrasensitive_2010}%
  \BibitemOpen
  \bibfield  {author} {\bibinfo {author} {\bibfnamefont {L.~T.}\ \bibnamefont
  {Hall}}, \bibinfo {author} {\bibfnamefont {C.~D.}\ \bibnamefont {Hill}},
  \bibinfo {author} {\bibfnamefont {J.~H.}\ \bibnamefont {Cole}},\ and\
  \bibinfo {author} {\bibfnamefont {L.~C.~L.}\ \bibnamefont {Hollenberg}},\
  }\href {https://doi.org/10.1103/PhysRevB.82.045208} {\bibfield  {journal}
  {\bibinfo  {journal} {Phys. Rev. B}\ }\textbf {\bibinfo {volume} {82}},\
  \bibinfo {pages} {045208} (\bibinfo {year} {2010})}\BibitemShut {NoStop}%
\bibitem [{\citenamefont {Pham}\ \emph {et~al.}(2012)\citenamefont {Pham},
  \citenamefont {{Bar-Gill}}, \citenamefont {Belthangady}, \citenamefont
  {Le~Sage}, \citenamefont {Cappellaro}, \citenamefont {Lukin}, \citenamefont
  {Yacoby},\ and\ \citenamefont {Walsworth}}]{pham_enhanced_2012}%
  \BibitemOpen
  \bibfield  {author} {\bibinfo {author} {\bibfnamefont {L.~M.}\ \bibnamefont
  {Pham}}, \bibinfo {author} {\bibfnamefont {N.}~\bibnamefont {{Bar-Gill}}},
  \bibinfo {author} {\bibfnamefont {C.}~\bibnamefont {Belthangady}}, \bibinfo
  {author} {\bibfnamefont {D.}~\bibnamefont {Le~Sage}}, \bibinfo {author}
  {\bibfnamefont {P.}~\bibnamefont {Cappellaro}}, \bibinfo {author}
  {\bibfnamefont {M.~D.}\ \bibnamefont {Lukin}}, \bibinfo {author}
  {\bibfnamefont {A.}~\bibnamefont {Yacoby}},\ and\ \bibinfo {author}
  {\bibfnamefont {R.~L.}\ \bibnamefont {Walsworth}},\ }\href
  {https://doi.org/10.1103/PhysRevB.86.045214} {\bibfield  {journal} {\bibinfo
  {journal} {Phys. Rev. B}\ }\textbf {\bibinfo {volume} {86}},\ \bibinfo
  {pages} {045214} (\bibinfo {year} {2012})}\BibitemShut {NoStop}%
\bibitem [{\citenamefont {Bishof}\ \emph {et~al.}(2013)\citenamefont {Bishof},
  \citenamefont {Zhang}, \citenamefont {Martin},\ and\ \citenamefont
  {Ye}}]{bishof_optical_2013}%
  \BibitemOpen
  \bibfield  {author} {\bibinfo {author} {\bibfnamefont {M.}~\bibnamefont
  {Bishof}}, \bibinfo {author} {\bibfnamefont {X.}~\bibnamefont {Zhang}},
  \bibinfo {author} {\bibfnamefont {M.~J.}\ \bibnamefont {Martin}},\ and\
  \bibinfo {author} {\bibfnamefont {J.}~\bibnamefont {Ye}},\ }\href
  {https://doi.org/10.1103/PhysRevLett.111.093604} {\bibfield  {journal}
  {\bibinfo  {journal} {Phys. Rev. Lett.}\ }\textbf {\bibinfo {volume} {111}},\
  \bibinfo {pages} {093604} (\bibinfo {year} {2013})}\BibitemShut {NoStop}%
\bibitem [{\citenamefont {Lange}\ \emph {et~al.}(2020)\citenamefont {Lange},
  \citenamefont {Huntemann}, \citenamefont {Sanner}, \citenamefont {Shao},
  \citenamefont {Lipphardt}, \citenamefont {Tamm},\ and\ \citenamefont
  {Peik}}]{lange_coherent_2020}%
  \BibitemOpen
  \bibfield  {author} {\bibinfo {author} {\bibfnamefont {R.}~\bibnamefont
  {Lange}}, \bibinfo {author} {\bibfnamefont {N.}~\bibnamefont {Huntemann}},
  \bibinfo {author} {\bibfnamefont {C.}~\bibnamefont {Sanner}}, \bibinfo
  {author} {\bibfnamefont {H.}~\bibnamefont {Shao}}, \bibinfo {author}
  {\bibfnamefont {B.}~\bibnamefont {Lipphardt}}, \bibinfo {author}
  {\bibfnamefont {{\relax Chr}.}~\bibnamefont {Tamm}},\ and\ \bibinfo {author}
  {\bibfnamefont {E.}~\bibnamefont {Peik}},\ }\href
  {https://doi.org/10.1103/PhysRevLett.125.143201} {\bibfield  {journal}
  {\bibinfo  {journal} {Phys. Rev. Lett.}\ }\textbf {\bibinfo {volume} {125}},\
  \bibinfo {pages} {143201} (\bibinfo {year} {2020})}\BibitemShut {NoStop}%
\bibitem [{\citenamefont {Shaniv}\ \emph {et~al.}(2019)\citenamefont {Shaniv},
  \citenamefont {Akerman}, \citenamefont {Manovitz}, \citenamefont {Shapira},\
  and\ \citenamefont {Ozeri}}]{shaniv_quadrupole_2019}%
  \BibitemOpen
  \bibfield  {author} {\bibinfo {author} {\bibfnamefont {R.}~\bibnamefont
  {Shaniv}}, \bibinfo {author} {\bibfnamefont {N.}~\bibnamefont {Akerman}},
  \bibinfo {author} {\bibfnamefont {T.}~\bibnamefont {Manovitz}}, \bibinfo
  {author} {\bibfnamefont {Y.}~\bibnamefont {Shapira}},\ and\ \bibinfo {author}
  {\bibfnamefont {R.}~\bibnamefont {Ozeri}},\ }\href
  {https://doi.org/10.1103/PhysRevLett.122.223204} {\bibfield  {journal}
  {\bibinfo  {journal} {Phys. Rev. Lett.}\ }\textbf {\bibinfo {volume} {122}},\
  \bibinfo {pages} {223204} (\bibinfo {year} {2019})}\BibitemShut {NoStop}%
\bibitem [{\citenamefont {Kaewuam}\ \emph {et~al.}(2020)\citenamefont
  {Kaewuam}, \citenamefont {Tan}, \citenamefont {Arnold}, \citenamefont
  {Chanu}, \citenamefont {Zhang},\ and\ \citenamefont
  {Barrett}}]{kaewuam_hyperfine_2020}%
  \BibitemOpen
  \bibfield  {author} {\bibinfo {author} {\bibfnamefont {R.}~\bibnamefont
  {Kaewuam}}, \bibinfo {author} {\bibfnamefont {T.~R.}\ \bibnamefont {Tan}},
  \bibinfo {author} {\bibfnamefont {K.~J.}\ \bibnamefont {Arnold}}, \bibinfo
  {author} {\bibfnamefont {S.~R.}\ \bibnamefont {Chanu}}, \bibinfo {author}
  {\bibfnamefont {Z.}~\bibnamefont {Zhang}},\ and\ \bibinfo {author}
  {\bibfnamefont {M.~D.}\ \bibnamefont {Barrett}},\ }\href
  {https://doi.org/10.1103/PhysRevLett.124.083202} {\bibfield  {journal}
  {\bibinfo  {journal} {Phys. Rev. Lett.}\ }\textbf {\bibinfo {volume} {124}},\
  \bibinfo {pages} {083202} (\bibinfo {year} {2020})}\BibitemShut {NoStop}%
\bibitem [{\citenamefont {Yeh}\ \emph {et~al.}(2023)\citenamefont {Yeh},
  \citenamefont {Grensemann}, \citenamefont {Dreissen}, \citenamefont
  {Fürst},\ and\ \citenamefont {Mehlstäubler}}]{yeh_robust_2023}%
  \BibitemOpen
  \bibfield  {author} {\bibinfo {author} {\bibfnamefont {C.-H.}\ \bibnamefont
  {Yeh}}, \bibinfo {author} {\bibfnamefont {K.~C.}\ \bibnamefont {Grensemann}},
  \bibinfo {author} {\bibfnamefont {L.~S.}\ \bibnamefont {Dreissen}}, \bibinfo
  {author} {\bibfnamefont {H.~A.}\ \bibnamefont {Fürst}},\ and\ \bibinfo
  {author} {\bibfnamefont {T.~E.}\ \bibnamefont {Mehlstäubler}},\ }\href
  {https://doi.org/10.1088/1367-2630/acfc14} {\bibfield  {journal} {\bibinfo
  {journal} {New J. Phys.}\ }\textbf {\bibinfo {volume} {25}},\ \bibinfo
  {pages} {093054} (\bibinfo {year} {2023})}\BibitemShut {NoStop}%
\bibitem [{\citenamefont {Dreissen}\ \emph {et~al.}(2022)\citenamefont
  {Dreissen}, \citenamefont {Yeh}, \citenamefont {F{\"u}rst}, \citenamefont
  {Grensemann},\ and\ \citenamefont
  {Mehlst{\"a}ubler}}]{dreissen_improved_2022}%
  \BibitemOpen
  \bibfield  {author} {\bibinfo {author} {\bibfnamefont {L.~S.}\ \bibnamefont
  {Dreissen}}, \bibinfo {author} {\bibfnamefont {C.-H.}\ \bibnamefont {Yeh}},
  \bibinfo {author} {\bibfnamefont {H.~A.}\ \bibnamefont {F{\"u}rst}}, \bibinfo
  {author} {\bibfnamefont {K.~C.}\ \bibnamefont {Grensemann}},\ and\ \bibinfo
  {author} {\bibfnamefont {T.~E.}\ \bibnamefont {Mehlst{\"a}ubler}},\ }\href
  {https://doi.org/10.1038/s41467-022-34818-0} {\bibfield  {journal} {\bibinfo
  {journal} {Nat. Commun.}\ }\textbf {\bibinfo {volume} {13}},\ \bibinfo
  {pages} {7314} (\bibinfo {year} {2022})}\BibitemShut {NoStop}%
\bibitem [{\citenamefont {Martínez-Lahuerta}\ \emph
  {et~al.}(2024)\citenamefont {Martínez-Lahuerta}, \citenamefont {Pelzer},
  \citenamefont {Dietze}, \citenamefont {Krinner}, \citenamefont {Schmidt},\
  and\ \citenamefont {Hammerer}}]{martinez-lahuerta_quadrupole_2023}%
  \BibitemOpen
  \bibfield  {author} {\bibinfo {author} {\bibfnamefont {V.~J.}\ \bibnamefont
  {Martínez-Lahuerta}}, \bibinfo {author} {\bibfnamefont {L.}~\bibnamefont
  {Pelzer}}, \bibinfo {author} {\bibfnamefont {K.}~\bibnamefont {Dietze}},
  \bibinfo {author} {\bibfnamefont {L.}~\bibnamefont {Krinner}}, \bibinfo
  {author} {\bibfnamefont {P.~O.}\ \bibnamefont {Schmidt}},\ and\ \bibinfo
  {author} {\bibfnamefont {K.}~\bibnamefont {Hammerer}},\ }\href
  {https://doi.org/10.1088/2058-9565/ad085b} {\bibfield  {journal} {\bibinfo
  {journal} {Quantum Sci. Technol.}\ }\textbf {\bibinfo {volume} {9}},\
  \bibinfo {pages} {015013} (\bibinfo {year} {2024})}\BibitemShut {NoStop}%
\bibitem [{\citenamefont {Pelzer}(2023)}]{pelzer_robust_2023}%
  \BibitemOpen
  \bibfield  {author} {\bibinfo {author} {\bibfnamefont {L.}~\bibnamefont
  {Pelzer}},\ }\emph {\bibinfo {title} {Robust Artificial Clock Transition by
  Continuous Dynamical Decoupling in Multi-Ion Calcium Crystals}},\ \href
  {https://doi.org/10.15488/13239} {Ph.D. thesis},\ \bibinfo  {school} {Leibniz
  Unversit\"at Hannover} (\bibinfo {year} {2023})\BibitemShut {NoStop}%
\bibitem [{\citenamefont {Hannig}\ \emph {et~al.}(2019)\citenamefont {Hannig},
  \citenamefont {Pelzer}, \citenamefont {Scharnhorst}, \citenamefont {Kramer},
  \citenamefont {Stepanova}, \citenamefont {Xu}, \citenamefont {Spethmann},
  \citenamefont {Leroux}, \citenamefont {Mehlst{\"a}ubler},\ and\ \citenamefont
  {Schmidt}}]{hannig_towards_2019}%
  \BibitemOpen
  \bibfield  {author} {\bibinfo {author} {\bibfnamefont {S.}~\bibnamefont
  {Hannig}}, \bibinfo {author} {\bibfnamefont {L.}~\bibnamefont {Pelzer}},
  \bibinfo {author} {\bibfnamefont {N.}~\bibnamefont {Scharnhorst}}, \bibinfo
  {author} {\bibfnamefont {J.}~\bibnamefont {Kramer}}, \bibinfo {author}
  {\bibfnamefont {M.}~\bibnamefont {Stepanova}}, \bibinfo {author}
  {\bibfnamefont {Z.~T.}\ \bibnamefont {Xu}}, \bibinfo {author} {\bibfnamefont
  {N.}~\bibnamefont {Spethmann}}, \bibinfo {author} {\bibfnamefont {I.~D.}\
  \bibnamefont {Leroux}}, \bibinfo {author} {\bibfnamefont {T.~E.}\
  \bibnamefont {Mehlst{\"a}ubler}},\ and\ \bibinfo {author} {\bibfnamefont
  {P.~O.}\ \bibnamefont {Schmidt}},\ }\href {https://doi.org/10.1063/1.5090583}
  {\bibfield  {journal} {\bibinfo  {journal} {Rev. Sci. Instrum.}\ }\textbf
  {\bibinfo {volume} {90}},\ \bibinfo {pages} {053204} (\bibinfo {year}
  {2019})}\BibitemShut {NoStop}%
\bibitem [{\citenamefont {Pyka}\ \emph {et~al.}(2014)\citenamefont {Pyka},
  \citenamefont {Herschbach}, \citenamefont {Keller},\ and\ \citenamefont
  {Mehlst{\"a}ubler}}]{pyka_high-precision_2014}%
  \BibitemOpen
  \bibfield  {author} {\bibinfo {author} {\bibfnamefont {K.}~\bibnamefont
  {Pyka}}, \bibinfo {author} {\bibfnamefont {N.}~\bibnamefont {Herschbach}},
  \bibinfo {author} {\bibfnamefont {J.}~\bibnamefont {Keller}},\ and\ \bibinfo
  {author} {\bibfnamefont {T.~E.}\ \bibnamefont {Mehlst{\"a}ubler}},\ }\href
  {https://doi.org/10.1007/s00340-013-5580-5} {\bibfield  {journal} {\bibinfo
  {journal} {Appl. Phys. B}\ }\textbf {\bibinfo {volume} {114}},\ \bibinfo
  {pages} {231} (\bibinfo {year} {2014})}\BibitemShut {NoStop}%
\bibitem [{\citenamefont {Beev}\ \emph {et~al.}(2017)\citenamefont {Beev},
  \citenamefont {Fenske}, \citenamefont {Hannig},\ and\ \citenamefont
  {Schmidt}}]{beev_low-drift_2017}%
  \BibitemOpen
  \bibfield  {author} {\bibinfo {author} {\bibfnamefont {N.}~\bibnamefont
  {Beev}}, \bibinfo {author} {\bibfnamefont {J.-A.}\ \bibnamefont {Fenske}},
  \bibinfo {author} {\bibfnamefont {S.}~\bibnamefont {Hannig}},\ and\ \bibinfo
  {author} {\bibfnamefont {P.~O.}\ \bibnamefont {Schmidt}},\ }\href
  {https://doi.org/10.1063/1.4983925} {\bibfield  {journal} {\bibinfo
  {journal} {Rev. Sci. Instrum.}\ }\textbf {\bibinfo {volume} {88}},\ \bibinfo
  {pages} {054704} (\bibinfo {year} {2017})}\BibitemShut {NoStop}%
\bibitem [{\citenamefont {Stenger}\ \emph {et~al.}(2002)\citenamefont
  {Stenger}, \citenamefont {Schnatz}, \citenamefont {Tamm},\ and\ \citenamefont
  {Telle}}]{stenger_ultraprecise_2002}%
  \BibitemOpen
  \bibfield  {author} {\bibinfo {author} {\bibfnamefont {J.}~\bibnamefont
  {Stenger}}, \bibinfo {author} {\bibfnamefont {H.}~\bibnamefont {Schnatz}},
  \bibinfo {author} {\bibfnamefont {C.}~\bibnamefont {Tamm}},\ and\ \bibinfo
  {author} {\bibfnamefont {H.}~\bibnamefont {Telle}},\ }\href
  {https://doi.org/10.1103/PhysRevLett.88.073601} {\bibfield  {journal}
  {\bibinfo  {journal} {Phys. Rev. Lett.}\ }\textbf {\bibinfo {volume} {88}},\
  \bibinfo {pages} {073601} (\bibinfo {year} {2002})}\BibitemShut {NoStop}%
\bibitem [{\citenamefont {Scharnhorst}\ \emph {et~al.}(2015)\citenamefont
  {Scharnhorst}, \citenamefont {W{\"u}bbena}, \citenamefont {Hannig},
  \citenamefont {Jakobsen}, \citenamefont {Kramer}, \citenamefont {Leroux},\
  and\ \citenamefont {Schmidt}}]{scharnhorst_high-bandwidth_2015}%
  \BibitemOpen
  \bibfield  {author} {\bibinfo {author} {\bibfnamefont {N.}~\bibnamefont
  {Scharnhorst}}, \bibinfo {author} {\bibfnamefont {J.~B.}\ \bibnamefont
  {W{\"u}bbena}}, \bibinfo {author} {\bibfnamefont {S.}~\bibnamefont {Hannig}},
  \bibinfo {author} {\bibfnamefont {K.}~\bibnamefont {Jakobsen}}, \bibinfo
  {author} {\bibfnamefont {J.}~\bibnamefont {Kramer}}, \bibinfo {author}
  {\bibfnamefont {I.~D.}\ \bibnamefont {Leroux}},\ and\ \bibinfo {author}
  {\bibfnamefont {P.~O.}\ \bibnamefont {Schmidt}},\ }\href
  {https://doi.org/10.1364/OE.23.019771} {\bibfield  {journal} {\bibinfo
  {journal} {Opt. Express}\ }\textbf {\bibinfo {volume} {23}},\ \bibinfo
  {pages} {19771} (\bibinfo {year} {2015})}\BibitemShut {NoStop}%
\bibitem [{\citenamefont {Matei}\ \emph {et~al.}(2017)\citenamefont {Matei},
  \citenamefont {Legero}, \citenamefont {H{\"a}fner}, \citenamefont {Grebing},
  \citenamefont {Weyrich}, \citenamefont {Zhang}, \citenamefont {Sonderhouse},
  \citenamefont {Robinson}, \citenamefont {Ye}, \citenamefont {Riehle},\ and\
  \citenamefont {Sterr}}]{matei_1.5_2017}%
  \BibitemOpen
  \bibfield  {author} {\bibinfo {author} {\bibfnamefont {D.~G.}\ \bibnamefont
  {Matei}}, \bibinfo {author} {\bibfnamefont {T.}~\bibnamefont {Legero}},
  \bibinfo {author} {\bibfnamefont {S.}~\bibnamefont {H{\"a}fner}}, \bibinfo
  {author} {\bibfnamefont {C.}~\bibnamefont {Grebing}}, \bibinfo {author}
  {\bibfnamefont {R.}~\bibnamefont {Weyrich}}, \bibinfo {author} {\bibfnamefont
  {W.}~\bibnamefont {Zhang}}, \bibinfo {author} {\bibfnamefont
  {L.}~\bibnamefont {Sonderhouse}}, \bibinfo {author} {\bibfnamefont {J.~M.}\
  \bibnamefont {Robinson}}, \bibinfo {author} {\bibfnamefont {J.}~\bibnamefont
  {Ye}}, \bibinfo {author} {\bibfnamefont {F.}~\bibnamefont {Riehle}},\ and\
  \bibinfo {author} {\bibfnamefont {U.}~\bibnamefont {Sterr}},\ }\href
  {https://doi.org/10.1103/PhysRevLett.118.263202} {\bibfield  {journal}
  {\bibinfo  {journal} {Phys. Rev. Lett.}\ }\textbf {\bibinfo {volume} {118}},\
  \bibinfo {pages} {263202} (\bibinfo {year} {2017})}\BibitemShut {NoStop}%
\bibitem [{\citenamefont {Kreuter}\ \emph {et~al.}(2005)\citenamefont
  {Kreuter}, \citenamefont {Becher}, \citenamefont {Lancaster}, \citenamefont
  {Mundt}, \citenamefont {Russo}, \citenamefont {H{\"a}ffner}, \citenamefont
  {Roos}, \citenamefont {H{\"a}nsel}, \citenamefont {{Schmidt-Kaler}},
  \citenamefont {Blatt},\ and\ \citenamefont
  {Safronova}}]{kreuter_experimental_2005}%
  \BibitemOpen
  \bibfield  {author} {\bibinfo {author} {\bibfnamefont {A.}~\bibnamefont
  {Kreuter}}, \bibinfo {author} {\bibfnamefont {C.}~\bibnamefont {Becher}},
  \bibinfo {author} {\bibfnamefont {G.~P.~T.}\ \bibnamefont {Lancaster}},
  \bibinfo {author} {\bibfnamefont {A.~B.}\ \bibnamefont {Mundt}}, \bibinfo
  {author} {\bibfnamefont {C.}~\bibnamefont {Russo}}, \bibinfo {author}
  {\bibfnamefont {H.}~\bibnamefont {H{\"a}ffner}}, \bibinfo {author}
  {\bibfnamefont {C.}~\bibnamefont {Roos}}, \bibinfo {author} {\bibfnamefont
  {W.}~\bibnamefont {H{\"a}nsel}}, \bibinfo {author} {\bibfnamefont
  {F.}~\bibnamefont {{Schmidt-Kaler}}}, \bibinfo {author} {\bibfnamefont
  {R.}~\bibnamefont {Blatt}},\ and\ \bibinfo {author} {\bibfnamefont {M.~S.}\
  \bibnamefont {Safronova}},\ }\href
  {https://doi.org/10.1103/PhysRevA.71.032504} {\bibfield  {journal} {\bibinfo
  {journal} {Phys. Rev. A}\ }\textbf {\bibinfo {volume} {71}},\ \bibinfo
  {pages} {032504} (\bibinfo {year} {2005})}\BibitemShut {NoStop}%
\bibitem [{\citenamefont {Monz}(2011)}]{monz_quantum_2011}%
  \BibitemOpen
  \bibfield  {author} {\bibinfo {author} {\bibfnamefont {T.}~\bibnamefont
  {Monz}},\ }\emph {\bibinfo {title} {Quantum Information Processing beyond Ten
  Ion-Qubits}},\ \href@noop {} {Ph.D. thesis},\ \bibinfo  {school} {University
  of Innsbruck} (\bibinfo {year} {2011})\BibitemShut {NoStop}%
\bibitem [{\citenamefont {Rosenband}\ \emph {et~al.}(2007)\citenamefont
  {Rosenband}, \citenamefont {Schmidt}, \citenamefont {Hume}, \citenamefont
  {Itano}, \citenamefont {Fortier}, \citenamefont {Stalnaker}, \citenamefont
  {Kim}, \citenamefont {Diddams}, \citenamefont {Koelemeij}, \citenamefont
  {Bergquist},\ and\ \citenamefont {Wineland}}]{rosenband_observation_2007}%
  \BibitemOpen
  \bibfield  {author} {\bibinfo {author} {\bibfnamefont {T.}~\bibnamefont
  {Rosenband}}, \bibinfo {author} {\bibfnamefont {P.~O.}\ \bibnamefont
  {Schmidt}}, \bibinfo {author} {\bibfnamefont {D.~B.}\ \bibnamefont {Hume}},
  \bibinfo {author} {\bibfnamefont {W.~M.}\ \bibnamefont {Itano}}, \bibinfo
  {author} {\bibfnamefont {T.~M.}\ \bibnamefont {Fortier}}, \bibinfo {author}
  {\bibfnamefont {J.~E.}\ \bibnamefont {Stalnaker}}, \bibinfo {author}
  {\bibfnamefont {K.}~\bibnamefont {Kim}}, \bibinfo {author} {\bibfnamefont
  {S.~A.}\ \bibnamefont {Diddams}}, \bibinfo {author} {\bibfnamefont
  {J.~C.~J.}\ \bibnamefont {Koelemeij}}, \bibinfo {author} {\bibfnamefont
  {J.~C.}\ \bibnamefont {Bergquist}},\ and\ \bibinfo {author} {\bibfnamefont
  {D.~J.}\ \bibnamefont {Wineland}},\ }\href
  {https://doi.org/10.1103/PhysRevLett.98.220801} {\bibfield  {journal}
  {\bibinfo  {journal} {Phys. Rev. Lett.}\ }\textbf {\bibinfo {volume} {98}},\
  \bibinfo {pages} {220801} (\bibinfo {year} {2007})}\BibitemShut {NoStop}%
\bibitem [{\citenamefont {Peik}\ \emph {et~al.}(1994)\citenamefont {Peik},
  \citenamefont {Hollemann},\ and\ \citenamefont {Walther}}]{peik_laser_1994}%
  \BibitemOpen
  \bibfield  {author} {\bibinfo {author} {\bibfnamefont {E.}~\bibnamefont
  {Peik}}, \bibinfo {author} {\bibfnamefont {G.}~\bibnamefont {Hollemann}},\
  and\ \bibinfo {author} {\bibfnamefont {H.}~\bibnamefont {Walther}},\ }\href
  {https://doi.org/10.1103/PhysRevA.49.402} {\bibfield  {journal} {\bibinfo
  {journal} {Phys. Rev. A}\ }\textbf {\bibinfo {volume} {49}},\ \bibinfo
  {pages} {402} (\bibinfo {year} {1994})}\BibitemShut {NoStop}%
\bibitem [{\citenamefont {Huang}\ \emph {et~al.}(2021)\citenamefont {Huang},
  \citenamefont {Zhang}, \citenamefont {Zeng}, \citenamefont {Zhang},
  \citenamefont {Hao}, \citenamefont {Chen}, \citenamefont {Wang},
  \citenamefont {Guan},\ and\ \citenamefont {Gao}}]{huang_nearly_2021}%
  \BibitemOpen
  \bibfield  {author} {\bibinfo {author} {\bibfnamefont {Y.}~\bibnamefont
  {Huang}}, \bibinfo {author} {\bibfnamefont {B.}~\bibnamefont {Zhang}},
  \bibinfo {author} {\bibfnamefont {M.}~\bibnamefont {Zeng}}, \bibinfo {author}
  {\bibfnamefont {H.}~\bibnamefont {Zhang}}, \bibinfo {author} {\bibfnamefont
  {Y.}~\bibnamefont {Hao}}, \bibinfo {author} {\bibfnamefont {Z.}~\bibnamefont
  {Chen}}, \bibinfo {author} {\bibfnamefont {M.}~\bibnamefont {Wang}}, \bibinfo
  {author} {\bibfnamefont {H.}~\bibnamefont {Guan}},\ and\ \bibinfo {author}
  {\bibfnamefont {K.}~\bibnamefont {Gao}},\ }\bibfield  {journal} {\bibinfo
  {journal} {preprint}\ }\href {https://doi.org/10.21203/rs.3.rs-120082/v1}
  {10.21203/rs.3.rs-120082/v1} (\bibinfo {year} {2021})\BibitemShut {NoStop}%
\bibitem [{\citenamefont {Hume}\ and\ \citenamefont
  {Leibrandt}(2016)}]{hume_probing_2016}%
  \BibitemOpen
  \bibfield  {author} {\bibinfo {author} {\bibfnamefont {D.~B.}\ \bibnamefont
  {Hume}}\ and\ \bibinfo {author} {\bibfnamefont {D.~R.}\ \bibnamefont
  {Leibrandt}},\ }\href {https://doi.org/10.1103/PhysRevA.93.032138} {\bibfield
   {journal} {\bibinfo  {journal} {Phys. Rev. A}\ }\textbf {\bibinfo {volume}
  {93}},\ \bibinfo {pages} {032138} (\bibinfo {year} {2016})}\BibitemShut
  {NoStop}%
\bibitem [{\citenamefont {D{\"o}rscher}\ \emph {et~al.}(2020)\citenamefont
  {D{\"o}rscher}, \citenamefont {{Al-Masoudi}}, \citenamefont {Bober},
  \citenamefont {Schwarz}, \citenamefont {Hobson}, \citenamefont {Sterr},\ and\
  \citenamefont {Lisdat}}]{dorscher_dynamical_2020}%
  \BibitemOpen
  \bibfield  {author} {\bibinfo {author} {\bibfnamefont {S.}~\bibnamefont
  {D{\"o}rscher}}, \bibinfo {author} {\bibfnamefont {A.}~\bibnamefont
  {{Al-Masoudi}}}, \bibinfo {author} {\bibfnamefont {M.}~\bibnamefont {Bober}},
  \bibinfo {author} {\bibfnamefont {R.}~\bibnamefont {Schwarz}}, \bibinfo
  {author} {\bibfnamefont {R.}~\bibnamefont {Hobson}}, \bibinfo {author}
  {\bibfnamefont {U.}~\bibnamefont {Sterr}},\ and\ \bibinfo {author}
  {\bibfnamefont {C.}~\bibnamefont {Lisdat}},\ }\href
  {https://doi.org/10.1038/s42005-020-00452-9} {\bibfield  {journal} {\bibinfo
  {journal} {Commun. Phys.}\ }\textbf {\bibinfo {volume} {3}},\ \bibinfo
  {pages} {1} (\bibinfo {year} {2020})}\BibitemShut {NoStop}%
\bibitem [{\citenamefont {Rosenband}\ and\ \citenamefont
  {Leibrandt}(2013)}]{rosenband_exponential_2013}%
  \BibitemOpen
  \bibfield  {author} {\bibinfo {author} {\bibfnamefont {T.}~\bibnamefont
  {Rosenband}}\ and\ \bibinfo {author} {\bibfnamefont {D.~R.}\ \bibnamefont
  {Leibrandt}},\ }\href@noop {} {\bibfield  {journal} {\bibinfo  {journal}
  {arXiv:1303.6357}\ } (\bibinfo {year} {2013})},\ \Eprint
  {https://arxiv.org/abs/1303.6357} {arxiv:1303.6357} \BibitemShut {NoStop}%
\bibitem [{\citenamefont {Borregaard}\ and\ \citenamefont
  {S{\o}rensen}(2013)}]{borregaard_efficient_2013}%
  \BibitemOpen
  \bibfield  {author} {\bibinfo {author} {\bibfnamefont {J.}~\bibnamefont
  {Borregaard}}\ and\ \bibinfo {author} {\bibfnamefont {A.~S.}\ \bibnamefont
  {S{\o}rensen}},\ }\href {https://doi.org/10.1103/PhysRevLett.111.090802}
  {\bibfield  {journal} {\bibinfo  {journal} {Phys. Rev. Lett.}\ }\textbf
  {\bibinfo {volume} {111}},\ \bibinfo {pages} {090802} (\bibinfo {year}
  {2013})}\BibitemShut {NoStop}%
\bibitem [{\citenamefont {Keller}\ \emph {et~al.}(2003)\citenamefont {Keller},
  \citenamefont {Lange}, \citenamefont {Hayasaka}, \citenamefont {Lange},\ and\
  \citenamefont {Walther}}]{keller_deterministic_2003}%
  \BibitemOpen
  \bibfield  {author} {\bibinfo {author} {\bibfnamefont {M.}~\bibnamefont
  {Keller}}, \bibinfo {author} {\bibfnamefont {B.}~\bibnamefont {Lange}},
  \bibinfo {author} {\bibfnamefont {K.}~\bibnamefont {Hayasaka}}, \bibinfo
  {author} {\bibfnamefont {W.}~\bibnamefont {Lange}},\ and\ \bibinfo {author}
  {\bibfnamefont {H.}~\bibnamefont {Walther}},\ }\href
  {https://doi.org/10.1007/s00340-003-1114-x} {\bibfield  {journal} {\bibinfo
  {journal} {Appl. Phys. B: Lasers Opt.}\ }\textbf {\bibinfo {volume} {76}},\
  \bibinfo {pages} {125} (\bibinfo {year} {2003})}\BibitemShut {NoStop}%
\end{thebibliography}%

\pagebreak
\newpage
\widetext
\begin{center}
\textbf{\large  Supplemental Material for: ``Multi-ion frequency reference using dynamical decoupling''}
\end{center}
\stepcounter{supequation}
\stepcounter{supfigure}
\makeatletter
\renewcommand{\theequation}{S\arabic{equation}}
\renewcommand{\thefigure}{S\arabic{figure}}
\renewcommand{\thetable}{S\arabic{table}}

\section{RF-Coils}
\label{subsec:transfer_function}
\begin{figure}[h]
\includegraphics[width=0.7\linewidth]{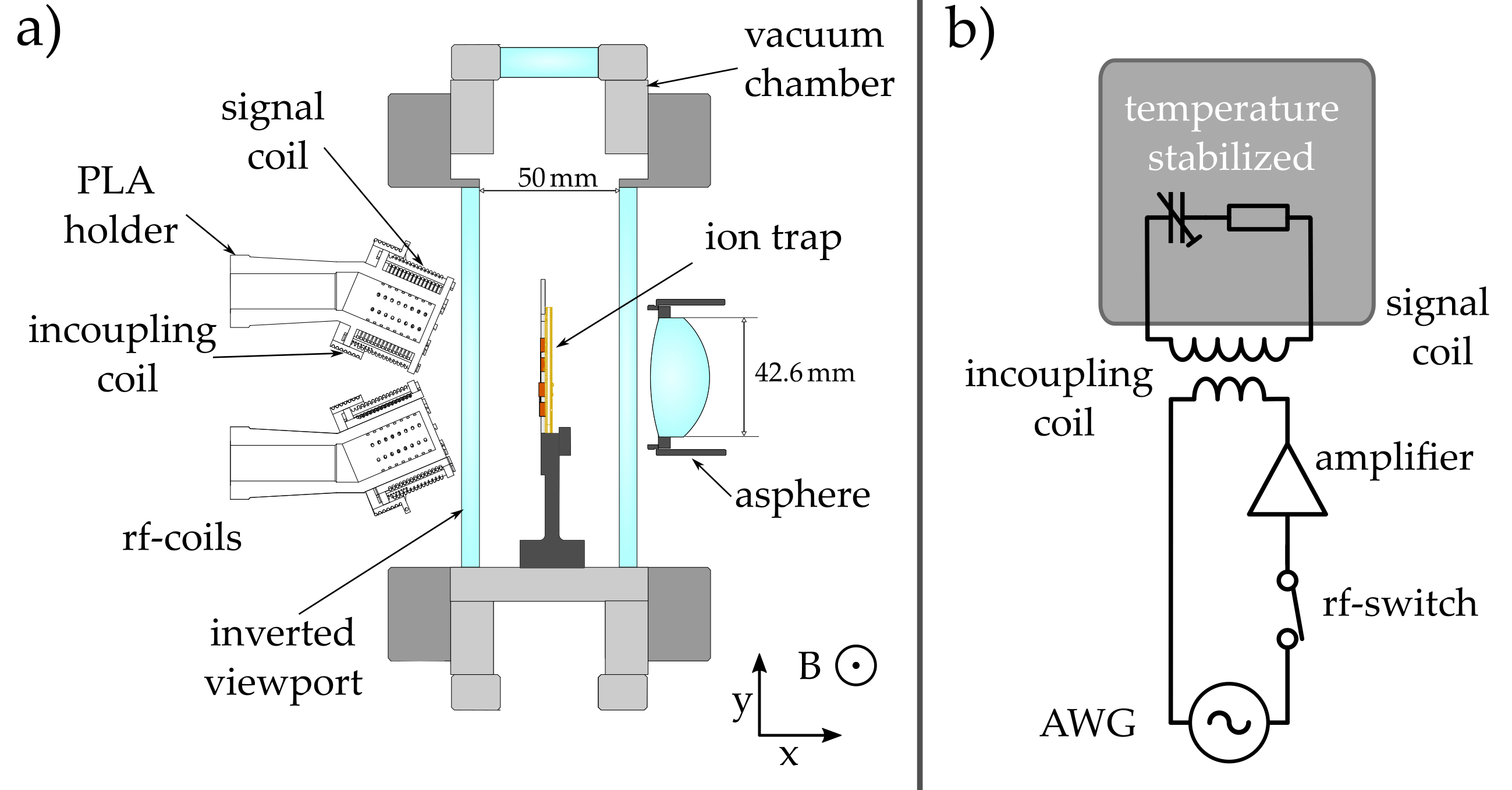}
\caption{a) Cross-section of the ion trap setup. The alignment of the rf-coils to the ion trap inside the vacuum vessel is displayed. b) Electric schematic of the rf-coil signal path. The alignment of the incoupling and signal coils is optimized for impedance matching between the \SI{50}{\ohm} input and the low-impedance signal coil.}
\label{fig:coil_setup}
\end{figure}

The design and placement of the rf-magnetic field coils for continuous dynamical decoupling (CDD) followed considerations as summarized in the following. The segmented Paul trap used for the experiments was designed to obtain low micromotion along the axial direction \cite{keller_deterministic_2003, hannig_towards_2019}. We therefore refrained from placing an rf-coil or wire close to the trapping zone, to avoid changing the trapping potential. Instead, the signal coils were placed outside the vacuum vessel inside an inverted viewport in a distance of approximately \SI{5}{\centi \meter} to the trap center (compare Fig.~\ref{fig:coil_setup}a). A resonant LCR circuit for signal amplification is necessary to achieve the required field strength. We choose quality factors of $Q_S = 8.59(3)$ and $Q_D =15.95(4)$. The signal distortion by the transfer function of the coils was taken into account in the applied waveform. For this, the transfer function $H(s)$ was measured with a receiver coil and modeled using
\begin{equation}\label{eq:transfer_function}
H(s) = g\frac{(s - z_r + iz_i)(s - z_r - iz_i)}{(s - p_r + ip_i)(s - p_r - ip_i)} \,,
\end{equation}
where the gain $g$, and a second order zero with real and imaginary values $z_r, z_i$ and a second order pole $p_r, p_i$ are chosen to fit the measured coil response.  The inverse of the transfer function was applied to the desired waveforms (Eqs.~\ref{eq:first_sweep} - \ref{eq:second_stage}) to pre-compensate the signal distortion. A two-channel arbitrary waveform generator\footnote{33622A, Keysight} (AWG) is used to generate the calculated waveform. An rf-switch, controlled by the experimental control system, ensures strict synchronization of the rf-pulses within the sequence. The signal is further amplified using an rf-amplifier\footnote{ZHL-3A, Mini-Circuits} (compare Fig.~\ref{fig:coil_setup}b).
With a typical input power ranging from \SIrange{20}{25}{\decibel}, the coils heat up by up to $\SI{10}{K}$, depending on the duty cycle of the rf-pulse.
The temperature dependence of the resonance frequency $f_0(T) =\frac{1}{\sqrt{L(T)C(T)}}$ of the circuit parts causes a variation of the coupling strength with altering duty cycle and changes in laboratory temperature. Therefore, the circuit design includes a temperature-controlled baseplate for the capacitive and resistive parts. The inductive part of the circuits consists of a copper coil held by a 3D printed polylactide (PLA) mount. For minimal heat build-up, the mount is constructed to provide efficient air cooling by convection.

\section{Waveforms}\label{sec_waveforms}
In the following, the employed waveforms for the CDD sequence are displayed. The required signal to obtain an interaction of the form given in Eq.~\ref{eq:drive} is displayed in Eq.~\ref{eq:second_stage}. If only the first stage CDD is used it simplifies to \ref{eq:first_drive}. Here, the Zeeman manifold index has been omitted for clarity. Eqs.~\ref{eq:first_sweep} and \ref{eq:second_sweep} are employed as preparation steps for first and second stage, respectively. Furthermore, the adiabatic transfer of the population is used for preparation of the desired dressed state. The waveforms are combined continuously by adapting the phases $\phi_i$ according to the preceding waveform. Note, that the signal amplitudes $A_1, A_2$ must be calibrated \textit{in situ} to match the desired coupling amplitudes $\Omega_1, \Omega_2$. The frequencies $\omega_i$, sweep durations $t_{\text{sw},i},\sigma$ and phases $\phi_i$ can be chosen with high precision. The frequency range of the frequency sweeps $\Delta \omega_{\text{sw},i} = \omega_{i} - \omega_{i,\text{init}}$ must be small compared to the energetic gap of the levels to avoid unwanted excitation. In addition, amplitude and frequency changes $\delta \Delta \omega_{\text{sw},i} /\delta t$ must be small enough to ensure adiabaticity. The preparation durations $t_{\text{sw},i}$ have been optimized \textit{in situ} for maximal peak excitation of the target state. The sign of the frequency sweep range $\Delta \omega_{\text{sw},i}$ determines which dressed ground state is populated. Using the parameters given in Table \ref{tab:CDD_parameter}, a target state population $>97\%$ is obtained.
\begin{align}
y(t) = & \frac{1}{2}e^{-\frac{(t - t_{\text{sw},1})^2}{\sigma^2}} A_1 \sin\left( \left[ \omega_{1,\text{init}} t + \frac{ (\omega_{1} - \omega_{1,\text{init}})}{ 2t_{\text{sw},1} }t^2 \right] \right)
\label{eq:first_sweep} \\ \eqname{First sweep} \\
y(t) =&  \frac{1}{2} A_1  \sin(t \omega_1 + \phi_{1}) \label{eq:first_drive}\\ \eqname{First stage} \\
y(t) =& \frac{1}{2} A_{1} \sin(\omega_1 t + \phi_{1} ) +  \frac{1}{2}e^{-\frac{(t - t_{\text{sw},2})^2}{\sigma^2}}  A_{2}  \sin(\omega_1 t + \frac{\pi}{2} + \phi_{1})\times \sin\left(\left[\omega_{2,\text{init}} t + \frac{ \omega_{2} - \omega_{2,\text{init}}}{2 t_{\text{sw},2}} t^2\right] + \phi_{2} \right)
\label{eq:second_sweep}\\ \eqname{Second sweep} \\
y(t) =& \frac{1}{2}  A_{1} \sin(\omega_1 t + \phi_{\text{fast}} ) +  \frac{1}{2}  A_{2}  \sin(\omega_1 t + \frac{\pi}{2} + \phi_{\text{fast}})\times \sin(\omega_{2} t + \phi_{\text{slow}} )
\label{eq:second_stage}\\ \eqname{Second stage}
\end{align}

\section{Continuous dynamical decoupling parameter}
\begin{table}[h]
\centering
\begin{tabular}{|c|r|r|}
\hline
Parameter Set &  Resonant & Magic mixing angle \\
\hline
\hline
$\omega_{S1}$ &  10002089& 10002089\\
\hline
$\omega_{D1}$ &  5994834& 5994834\\
\hline
$\Omega_{S1}$ &  46862& 46862\\
\hline
$\Omega_{D1}$ &  115446& 115446\\
\hline
$\Omega_{S2}$ &  3469  & 3469\\
\hline
$\Omega_{D2}$ &  6809& 6809\\
\hline
$\omega_{S2}$  & 46915 & 46951\\
\hline
$\omega_{D2}$  & 69287 & 70731\\
\hline
$\Delta \omega_{\text{sw1}}$  & -150000 & -1500000\\
\hline
$t_{\text{sw},1}$  & 500& 500\\
\hline
$\Delta \omega_{\text{sw2}}$  & 80000& 30000\\
\hline
$t_{\text{sw},2}$ & 7000& 7000\\
\hline

\end{tabular}

\caption{Parameter sets for CDD waveforms. All units of frequencies $\omega_i$, coupling strength $\Omega_i$ and sweep range $\Delta \omega_{\text{sw},i}$ are in  \unit{\hertz}, all units for sweep times $t_{\text{sw},i}$ are given in  $\unit{\micro \second}$.  }
\label{tab:CDD_parameter}
\end{table}

\end{document}